\documentclass[apjl]{emulateapj}
\usepackage{amsfonts,amsmath,graphicx,natbib,apjfonts,subfigure,color,gensymb,longtable}
\usepackage[breaklinks,colorlinks, urlcolor=blue,citecolor=blue,linkcolor=blue]{hyperref}

\def\nu{1}
\def\ou{2}
\def\col{3}
\def\fer{4}
\def\cal{8}
\def\cfa{6}
\def\nyu{7}

\def\carnegie{9}
\def\hubble{10}
\def\fr{5}

\shorttitle{X-rays from SLSNe}
\shortauthors{Margutti }

\begin{document}
\title{Results from a systematic survey of X-ray emission from Hydrogen-poor Superluminous SNe}

\author{R. Margutti\altaffilmark{\nu}, 
R. Chornock\altaffilmark{\ou}, 
B. D. Metzger\altaffilmark{\col}, 
D. L. Coppejans\altaffilmark{\nu}, 
C. Guidorzi\altaffilmark{\fer}, 
G. Migliori\altaffilmark{\fr},
D. Milisavljevic\altaffilmark{\cfa}, 
E. Berger\altaffilmark{\cfa}, 
M. Nicholl\altaffilmark{\cfa}, 
B.~A. Zauderer\altaffilmark{\nyu}, 
%W. Fong\altaffilmark{\az},  
R. Lunnan\altaffilmark{\cal}, 
%J. T. Parrent\altaffilmark{\cfa}, 
A. Kamble\altaffilmark{\cfa}, 
M. Drout\altaffilmark{\carnegie,\hubble}, 
M. Modjaz\altaffilmark{\nyu}
%A. ~M. Soderberg\altaffilmark{\cfa}, 
%M. Sullivan\altaffilmark{\st}
 %F. Bianco\altaffilmark{\nyu}, L. Chomiuk\altaffilmark{\msu}, ,  Y. Liu\altaffilmark{\nyu},A. Mac Fadyen\altaffilmark{\nyu} }
}

\altaffiltext{\nu}{Center for Interdisciplinary Exploration and Research in Astrophysics (CIERA) and Department of Physics and Astronomy, Northwestern University, Evanston, IL 60208}
\altaffiltext{\ou}{Astrophysical Institute, Department of Physics and Astronomy, 251B Clippinger Lab, Ohio University, Athens, OH 45701, USA}
\altaffiltext{\col}{Columbia Astrophysics Laboratory, Columbia University, Pupin Hall, New York, NY 10027, USA}
\altaffiltext{\fer}{Department of Physics and Earth Science, University of Ferrara, via Saragat 1, I-44122, Ferrara, Italy}
\altaffiltext{\fr}{Laboratoire  AIM  (CEA/IRFU  -  CNRS/INSU  -  Universite  Paris  Diderot),  CEA  DSM/SAp,  F-91191  Gif-sur-Yvette,
France}
\altaffiltext{\cfa}{Harvard-Smithsonian Center for Astrophysics, 60 Garden St., Cambridge, MA 02138, USA}
\altaffiltext{\nyu}{Center for Cosmology and Particle Physics, New York University, 4 Washington Place, New York, NY 10003, USA}
\altaffiltext{\cal}{Department of Astronomy, California Institute of Technology, 1200 East California Boulevard, Pasadena, CA 91125, USA}
\altaffiltext{\carnegie}{Carnegie Observatories, 813 Santa Barbara Street, Pasadena, CA 91101, USA 2}
%\altaffiltext{\az}{Steward Observatory, University of Arizona, 933 N. Cherry Avenue, Tucson, AZ 85721, USA}
\altaffiltext{\hubble}{Hubble, Carnegie-Dunlap Fellow}
%\altaffiltext{\msu}{Department of Physics and Astronomy, Michigan State University, East Lansing, MI:48824, USA}
%\altaffiltext{\st}{Department of Physics and Astronomy, University of Southampton, Southampton SO17 1BJ, UK}

\begin{abstract}
We present the results from a sensitive X-ray survey of 26 nearby hydrogen-poor superluminous supernovae (SLSNe-I) with \emph{Swift}, Chandra and XMM.  This dataset constrains the SLSN evolution from a few days until $\sim2000$ days after explosion, reaching a luminosity $L_x\sim10^{40}\,\rm{erg\,s^{-1}}$ and revealing the presence of  significant X-ray emission at the location of PTF12dam. No SLSN-I is detected above $L_x\sim10^{41}\,\rm{erg\,s^{-1}}$, suggesting that the luminous X-ray emission $L_x\sim10^{45}\,\rm{erg\,s^{-1}}$ associated with SCP60F6 is not common among SLSNe-I. We  constrain the presence of off-axis GRB jets, ionization breakouts from magnetar central engines and the density in the sub-pc environments of SLSNe-I through Inverse Compton emission. The deepest limits rule out  the weakest uncollimated GRB outflows, suggesting that \emph{if} the similarity of SLSNe-I with GRB/SNe extends to their fastest ejecta, then 
SLSNe-I are either powered by energetic jets pointed far away from our line of sight ($\theta>30\degree$), or harbor failed jets that do not successfully break through the stellar envelope. Furthermore, \emph{if} a magnetar central engine is responsible for the exceptional luminosity of SLSNe-I, our X-ray analysis favors large magnetic fields $B>2\times10^{14}$ G and ejecta masses $M_{ej}>3\,\rm{M_{\sun}}$ in agreement with optical/UV studies. Finally, we constrain the pre-explosion mass-loss rate of stellar progenitors of SLSNe-I. For PTF12dam we infer $\dot M<2\times10^{-5}\,\rm{M_{\sun}yr^{-1}}$, suggesting that the SN shock interaction with the CSM is unlikely to supply the main source of energy powering the optical transient and that some SLSN-I progenitors end their life as compact stars surrounded by a low-density medium similar to long GRBs and Type Ib/c SNe.
\end{abstract}
%%%%%%%%%%%%%%%%%%%%%%%%%%%%%%%%%%%%%%%%%%%
\section{Introduction}

Superluminous supernovae (SLSNe) are among the most luminous known stellar explosions in the Universe.
Recognized as a class only in 2009 (\citealt{Quimby11}, \citealt{Chomiuk11}), SLSNe reach optical-UV luminosities $L$$>$$7\times10^{43}\,\rm{erg\,s^{-1}}$, $\sim10-100$ more luminous than common SNe, and are likely associated with the death of massive stars.
The source of energy powering their exceptional energy release is still debated (e.g. \citealt{Gal-Yam12}).  
The proposed energy sources include: (i) radioactive decay of large amounts of freshly synthesized $^{56}$Ni  ($M_{Ni}$$\gtrsim$$ 5\,M_{\odot}$), a signature of pair-instability explosions (as proposed for SN\,2007bi, \citealt{Gal-Yam09}); (ii) SN shock interaction with dense material in the environment (e.g. \citealt{Smith07}, \citealt{Chevalier11}); (iii) a magnetar central engine (e.g. \citealt{Kasen10,Woosley10,Nicholl13}).  The narrow features ($v\lesssim100\,\rm{km\,s^{-1}}$)  in the spectra of hydrogen-rich SLSNe like SN2006gy clearly indicate that the interaction of the SN blast wave with the medium plays a role (e.g. \citealt{Smith07,Ofek07}), while it is unclear if a single mechanism can power  hydrogen-stripped SLSNe (i.e. SLSNe-I). Indeed, SLSN-I iPTF13ehe has been interpreted as the combination of energy extracted from a magnetar central engine coupled with radiation from the radioactive decay of $\sim2.5\rm{M_{\sun}}$ of $^{56}$Ni, and a late-time interaction of the SN shock with the medium (\citealt{Yan15,Wang16}).

A number of independent lines of evidence support the idea that SLSNe-I might harbor an engine.
Observations of SLSN-I host galaxies indicate a preference for low-metallicity environments, which inspired a connection with  long Gamma-Ray Bursts (GRBs, \citealt{Lunnan14,Leloudas15b,Perley16,Chen16}, see however \citealt{Angus16}). Along the same line, \cite{Greiner15} reported the detection of typical SLSN spectral features in SN2011kl associated with GRB\,111209A, and suggested that a magnetar central engine powered both the initial burst of $\gamma$-rays and the later optical/UV SN emission (\citealt{Metzger15}).  
\citet{Milisavljevic13} found links between the late-time emission properties of a subset of energetic, slow-evolving supernovae and the superluminous SN\,2007bi. They suggested that a single, possibly jetted, explosion mechanism may unify all of these events that span $-21 \le M_B \le -17$ mag. 
Additionally, nebular spectroscopic studies by \cite{Nicholl16} revealed similarities between the SLSN-I 2015bn and SN\,1998bw, associated with GRB\,980425, suggesting that the cores of their massive progenitors shared a similar structure at the time of collapse. 
In another source, SLSN-I Gaia 16apd,  \cite{Nicholl17} further demonstrate that the luminous excess of UV emission originates from a central source of energy, rather then reduced UV absorption or shock interaction with a thick medium (see however \citealt{Yan16}). Finally, luminous X-ray emission has been detected at the location of the SLSN-I SCP06F6 \citep{Gansicke09} with luminosity $L_x\sim 10^{45}\,\rm{erg\,s^{-1}}$ $\sim$$70$ days (rest-frame) after the explosion (\citealt{Levan13}).  At this epoch, SCP06F6 even outshines GRBs by a large factor, suggesting the presence of a still-active central engine that manifests itself through very luminous and long-lasting X-ray emission \citep{Levan13,Metzger15}. Before our effort, SCP06F6 was the only SLSN-I for which an X-ray source was detected at a location consistent with the optical transient. 

These observational results suggest a connection between SLSNe-I and engine-driven SNe. However, it is not yet known how the properties of the engines (successful jet? relativistic ejecta? collimated or spherical central-engine powered outflow?), progenitor stars, and circumstellar environments would compare.
Here we present the results from a systematic search for X-ray emission from SLSNe-I both at early and at late times, which directly depends on the properties of the immediate environment and central engine (if any). The direct detection of the stellar progenitors of SLSNe-I in  pre-explosion optical images is not possible due to their large distances ($z\geq0.1$). Sampling the circumstellar density profile in the closest environment is thus the most direct probe of their progenitors and their recent mass-loss history before stellar death.

The dataset that we present here includes the deepest X-ray observations of SLSNe-I with \emph{Swift}, XMM and the Chandra X-ray Observatory (CXO), extending from the time of discovery until $\sim$2000 days (rest-frame) after explosion and led to the discovery of X-ray emission at the location of the slowly evolving SLSN-I PTF12dam. 
These observations, described in Sec. \ref{Sec:Obs}, indicate that superluminous X-ray emission similar to what was observed in association with SCP06F6 is not common in SLSNe-I (Sec. \ref{Sec:SLXray}) and allow us to place  meaningful constraints on the environment density at the SLSN site (Sec. \ref{Sec:IC}).  We constrain the properties of central engines in SLSNe-I in Sec. \ref{Sec:CE} by investigating the presence of late-time X-ray re-brightenings that can either be due to emission from off-axis collimated relativistic outflows similar to GRBs,  or to the ionization breakouts from magnetar central engines \citep{Metzger14}. %(Sec. \ref{Sec:GRB}, \ref{Sec:magnetar}). 
Conclusions are drawn in Sec. \ref{Sec:Conc}.

%%%%%%%%%%%%%%%%%%%%%%%%%%%%%%%%%%%%%%%%%%
\section{X-ray observations and analysis}
\label{Sec:Obs}

\begin{deluxetable*}{lcccccc}[h!]
\tabletypesize{\scriptsize}
%\rotate
\tablecolumns{4} 
\tablewidth{40pc}
\tablecaption{Gold Sample.}
\tablehead{  \colhead{SN} & \colhead{z} & \colhead{$d_{L}$}&   Discovery Date&  Inferred   Explosion Date &\colhead{$\rm{NH}_{\rm{MW}}$ }& \colhead{Instrument}\\
 & & (Mpc) & (MJD)&(MJD)&($10^{20}\rm{cm^{-2}}$) &}
\startdata
SCP06F6 & 1.189& 8310& 53787\footnote{From \cite{Levan13}.} &53767\footnote{The time of peak is MJD 53872. The rise time is $\sim50$ days in the rest frame  \citep{Barbary09}. }&$0.885$  & XMM+CXO \\
PTF12dam & 0.107 &  498 &56037$^{\rm{a}}$&56022\footnote{The light-curve reached maximum light on MJD 56088 and the rest-frame rise-time is $\sim60$ days \citep{Nicholl13}. }& $1.11$ & Swift+CXO\\
PS1-14bj & 0.521 & 3012 &56618\footnote{From \cite{Lunnan16}.}&56611\footnote{\cite{Lunnan16} estimate a peak time on MJD 56801.3 and a rest-frame rise-time $\gtrsim 125$ days. }& $1.71$& XMM\\
SN\,2015bn/PS15ae &  0.1136& 513.2 &57014\footnote{From \cite{Nicholl16}.}&57013\footnote{The SN reached $r-$ band maximum light on MJD 57102 \citep{Nicholl16}. The rise-time inferred by \cite{Nicholl16} is $\sim80$ days in the  rest-frame.}&$2.37$& Swift+XMM\\
%ASAS-SN15lh&0.2326 & 1171&$3.07\times 10^{20}$ &Swift+Chandra \\
\enddata
\label{Tab:Gold}
\end{deluxetable*}

\begin{deluxetable*}{lcccccc}[h!]
\tabletypesize{\scriptsize}
%\rotate
\tablecolumns{4} 
\tablewidth{40pc}
\tablecaption{Bronze Sample.}
\tablehead{  \colhead{SN} & \colhead{z} & \colhead{$d_{L}$}& Discovery Date& Inferred Explosion Date &\colhead{$\rm{NH}_{\rm{MW}}$ }& \colhead{Instrument}\\
 & & (Mpc) & (MJD) & (MJD) &($10^{20}\rm{cm^{-2}}$) &}
\startdata
PTF09cnd & 0.258 &	1317 &55025\footnote{From \citep{Levan13}.}&55006\footnote{From \cite{Quimby11}, the peak time is MJD 55069.145 and the rest-frame rise-time is $\sim50$ days.}&$2.20$ & Swift+XMM\\
SN\,2010gx & 0.230 &	1156 &55260$^{\rm{a}}$&55251\footnote{From \cite{Quimby11}, the peak time is MJD 55279 and the rest-frame rise-time is $\sim23$ days.}& $3.28$ & Swift\\
SN\,2010kd & 0.101&	 468 &55453$^{\rm{a}}$&55398\footnote{\cite{Vinko12} report that SN2010kd reached maximum light 40 days after discovery. We assume a 50 day rest-frame rise-time.  }& $2.32$ & Swift\\
SN\,2011ke & 0.143&	 682 &55650$^{\rm{a}}$&55649\footnote{From \cite{Inserra13}.}&$1.27$ & Swift+CXO\\
SN\,2012il & 0.175&	 851	  &55926$^{\rm{a}}$&55919$^{\rm{e}}$&$2.38$ & Swift\\
iPTF13ehe & 0.3434 & 1833 &56621\footnote{From \cite{Yan15}.}&56496.4\footnote{\cite{Yan15} report a range of explosion dates between MJD 56470.8 and MJD 56522.0. We use the middle date MJD 56496.4.}& $4.30$ & Swift\\
LSQ14mo & 0.253 & 1288 &56687\footnote{From \cite{Leloudas15}.}&56624\footnote{Peak time on MJD 56699 \citep{Leloudas15}. The pre-max evolution is only sparsely sampled (see \citealt{Leloudas15}). We assume a 50-day rest-frame rise-time, similar to other SLSNe-I.}& $6.59$ & Swift\\
LSQ14an & 0.163 & 787 &56689\footnote{From \cite{Jerkstrand16}.}&56513$^{\rm{j}}$& $6.13$ & Swift\\
CSS140925-005854 & 0.46& 2590&56920\footnote{From  the CRTS source catalog http://nesssi.cacr.caltech.edu/catalina/AllSN.html}&56900$^{\rm{k}}$&  $3.99$ & Swift\\
LSQ14fxj & 0.36 & 1937 &56942\footnote{From \cite{SmithM14}.}&56872\footnote{According to \cite{SmithM14}, on Nov 22, 2014 the transient was 4-5 weeks rest-frame after maximum light. The inferred time of maximum light is MJD 56940. We assume a 50-day rest-frame rise-time. }& $3.28$ & Swift\\
%PS15br & 0.101 & 468 & $4.33\times 10^{20}$ & Swift\\
DES15S2nr & 0.220 & 1099 &57251\footnote{From \cite{DAndrea15}.}&57251\footnote{Very sparse photometric coverage. On MJD 57286 \cite{DAndrea15} report that the transient is still before peak. We adopt the discovery date as a rough proxy for the explosion date here.}& $3.02$ & Swift\\
SN\,2016ard/PS16aqv & 0.2025\footnote{From Blanchard et al., in prep.} & 988 &57438\footnote{From http://star.pst.qub.ac.uk/ps1threepi/psdb/public/}&57393\footnote{The peak time is MJD 57453 from http://star.pst.qub.ac.uk/ps1threepi/psdb/public/  We assume a 50-day rest-frame rise time.}&$3.97$ & Swift\\
\enddata
\label{Tab:Bronze}
\end{deluxetable*}

\begin{deluxetable*}{lcccccc}[h!]
\tabletypesize{\scriptsize}
%\rotate
\tablecolumns{4} 
\tablewidth{40pc}
\tablecaption{Iron Sample. }
\tablehead{  \colhead{SN} & \colhead{z} & \colhead{$d_{L}$}&   Discovery Date& Inferred Explosion Date& \colhead{$\rm{NH}_{\rm{MW}}$ }& \colhead{Instrument}\\
 & & (Mpc) &(MJD)& (MJD)&($10^{20}\rm{cm^{-2}}$) &}
\startdata
SN2009jh/PTF09cwl& 0.349 &1868 &55010\footnote{From \cite{Levan13}.}&55010\footnote{The time of peak is MJD 55081 and the rise time is $\sim50$ days in the rest frame \citep{Quimby11}.}& $1.49$ & Swift\\
PTF09atu& 0.501 & 2870& 55016$^{\rm{a}}$&54988\footnote{The time of maximum light is MJD 55063 \citep{Quimby11}. We assume a 50-day rise-time in the rest-frame.}& $3.79$ & Swift\\
PTF10aagc  & 0.207 &1027 &Unclear&55413\footnote{The time of maximum light is MJD 55473 \citep{Perley16}. We assume a 50 day rise-time in the rest frame.}& $2.61$ & Swift\\
SN\,2010md/PTF10hgi& 0.098 & 463 &55331$^{\rm{a}}$&55323\footnote{From \cite{Inserra13}. }& $5.81$ & Swift\\
PS1-11bdn & 	0.738	& 4601&55910.4\footnote{From Lunnan et al., in prep.}&55889.2$^{\rm{f}}$& $3.76$ & Swift\\
PTF11rks  & 0.19&  	 933 &55916$^{\rm{a}}$&55912$^{\rm{e}}$& $4.66$ & Swift\\
DES15C3hav & 0.392 & 2142&57310\footnote{From \cite{Challis16}.}&57270\footnote{From \cite{Challis16} the peak time is MJD 57340. We assume a 50 day rest-frame rise-time.}& $0.705$ & Swift\\
OGLE15qz & 0.63 & 3790&57264&57264\footnote{From http://ogle.astrouw.edu.pl/ogle4/transients/transients.html}&$4.28$ & Swift\\
OGLE15sd & 0.656     & 3319          &57295&57295$^{\rm{i}}$& $9.44$ & Swift\\
PS16op  & 0.48 & 2726&57398\footnote{From \cite{Dimitriadis16}.}&57323\footnote{The peak time is MJD 57397 \citep{Dimitriadis16}. We assume a 50 day rest-frame rise-time.}& $6.73$ & Swift\\
\enddata
\label{Tab:Iron}
\end{deluxetable*}

\begin{deluxetable*}{lcccccc}[h!]
\tabletypesize{\scriptsize}
%\rotate
\tablecolumns{4} 
\tablewidth{35pc}
\tablecaption{Magnetar Parameters (magnetic field $B$, spin period $P_i$ and ejecta mass $M_{ej}$), estimated from the bolometric optical emission, and corresponding ionization break out times $t_{ion}$ and X-ray luminosities $L_{x}(t_{ion})$}
\tablehead{  \colhead{SN} & \colhead{$B$} & \colhead{$P_i$}& \colhead{$M_{ej}$} & Ref & $t_{ion}$ & $L_{x}(t_{ion})$\\
 & (G) & (ms) & ($M_{\odot}$) & & (yr) & $(\rm{erg\,s^{-1}})$ }
\startdata
SN\,2010md/PTF10hgi &$3.6\times10^{14}$&7.2& 3.9 &\cite{Inserra13} & 76.3 & $7.8\times10^{36}$\\
SN\,2010gx &$7.4\times10^{14}$& 2.0 & 7.1& \cite{Inserra13}& 1070 & $9.4\times10^{33}$ \\
PTF11rks &$6.8\times10^{14}$&7.50 & 2.8& \cite{Inserra13}&140&$6.5\times10^{35}$  \\
SN\,2011ke &$6.4\times10^{14}$& 1.7 & 8.6& \cite{Inserra13}&1170& $1.0\times10^{34}$ \\
PTF12dam & $5\times10^{13}$ & 2.3 & $7$ & \cite{Nicholl13}&4.7&$1.0\times10^{41}$\\
SN\,2012il & $4.1\times10^{14}$& 6.1 & 2.3 & \cite{Inserra13}&34.5&  $3.0\times10^{37}$\\
iPTF13ehe & $8\times10^{13}$& 2.55 & 35 &  \cite{Wang15}&304&$1.0\times10^{37}$\\
PS1-14bj & $10^{14}$ & 3.1 & 22.5& \cite{Lunnan16}&196& $1.5\times10^{37}$\\
                & $5\times10^{13}$ & 3.1 & 16& \cite{Lunnan16} w. leakage & 24.8&$3.8\times10^{39}$\\
SN\,2015bn/PS15ae   &  $0.9\times10^{14}$ & 2.1 & 8.4 & \cite{Nicholl16} &22.1&$1.5\times10^{39}$ \\   
		& $10^{14}$ & 1.7 & 15.1 & \cite{Nicholl16} &88.4&$7.6\times10^{37}$\\
		&$0.9\times10^{14}$&	2.1 &	8.3	&  \cite{Nicholl16b} & 21.6& $1.6\times10^{39}$\\
		&$0.2\times10^{14}$&1.5	&	7.4	&  \cite{Nicholl16b} & 0.64&$3.5\times10^{43}$ \\
		&$0.9\times10^{14}$&2.2	&	11.9	&  \cite{Nicholl16b} & 44.5& $3.7\times10^{38}$ \\
		&$0.4\times10^{14}$&	1.8&	9.0	&  \cite{Nicholl16b} &5.02& $1.5\times10^{41}$  \\
\enddata
\label{Tab:Magnetar}
\end{deluxetable*}

\begin{figure}
\includegraphics[scale=0.6]{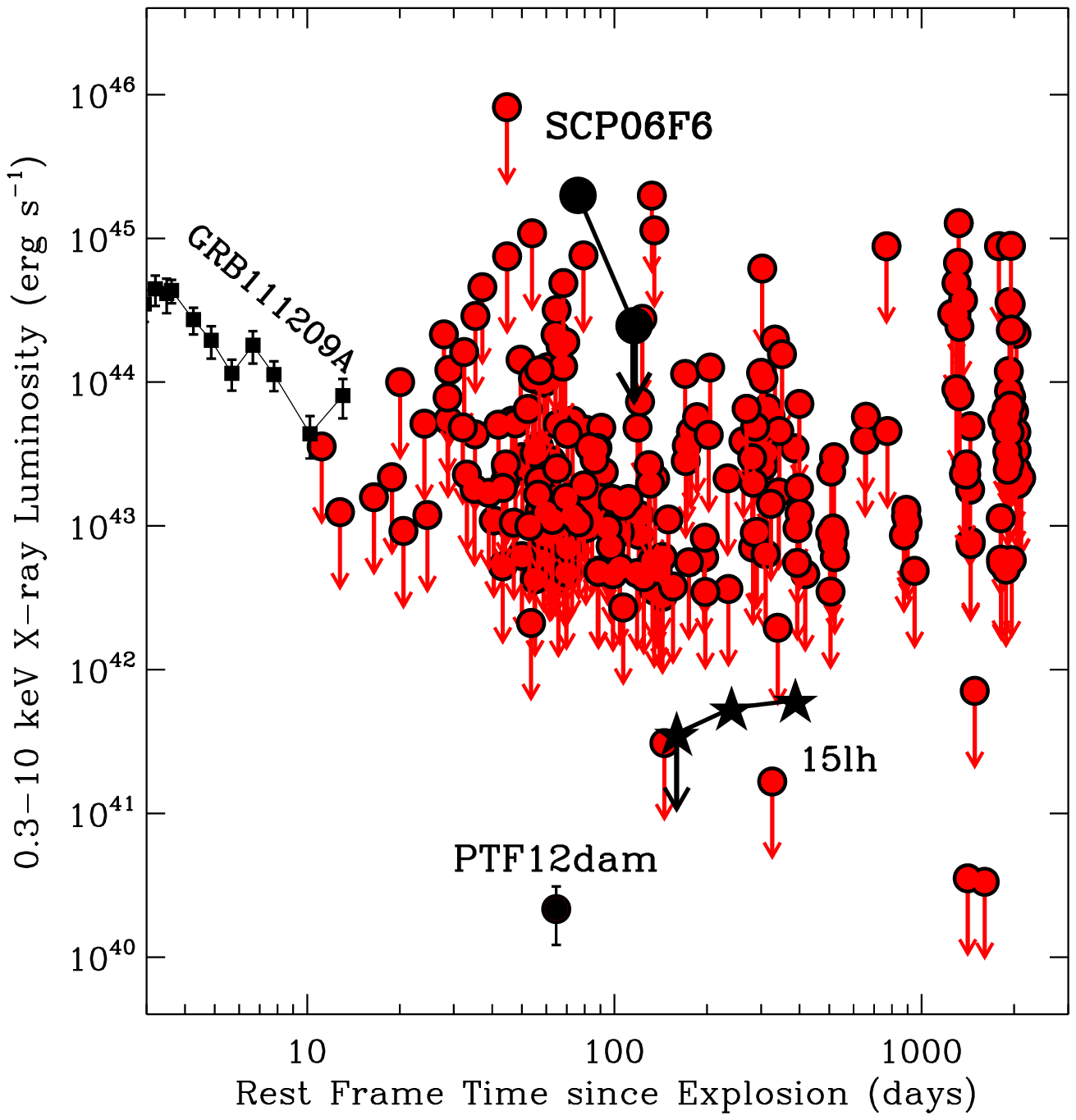}
\caption{X-ray observations of SLSNe-I spanning the time range $\sim$10-2000 days (red circles for upper limits, black circles for detections) show that superluminous X-ray emission of the kind detected at the location of SCP06F6 (\citealt{Gansicke09}, \citealt{Levan13}) is not common. Black stars: X-ray emission at the location of ASASSN-15lh \citep{Margutti16}, which has a very disputed physical origin. Black squares: X-ray afterglow of GRB\,112109A, associated with the over-luminous SN 2011kl \citep{Greiner15}.}
\label{Fig:ddd}
\end{figure}

%%%%%%%%%%%%%%%%%%%%%%%%%%%%%%%%%%%%%%%%%%%
Since 2011, we routinely followed up all publicly announced nearby ($z\lesssim0.5$) SLSNe-I with \emph{Swift}-XRT with a series of observations acquired between the time of discovery and $\sim360$ days (rest-frame) after explosion. For a subset of events we acquired deep X-ray observations with dedicated programs on the  Chandra X-ray Observatory (CXO) and XMM-\emph{Newton}. Additionally, we searched the  \emph{Swift}-XRT, CXO and XMM archives for serendipitous or unpublished observations of SLSNe-I discovered before May 2016. 
Our final sample consists of %We collected deep \emph{Swift}-XRT,  Chandra X-ray Observatory (CXO) and XMM-\emph{Newton} X-ray observations of 
%collected the largest sample of X-ray observations of nearby hydrogen-poor SLSNe. Our sample consists 
 26 SLSNe-I discovered between 2006 and May 2016. The dataset covers the time range between $\sim$days after explosion until $\sim2000$ days (rest-frame), and comprises $\sim$ 700 hrs of observations.  We update the X-ray observations of the sample of 11 SLSNe-I from \cite{Levan13} with the most recent data\footnote{Note that PTF11dsf and CSS121015, included by \cite{Levan13} in the sample of SLSNe-I are in fact H-rich events (see \citealt{Benetti14, Perley16}). Additionally, for  PTF11dsf an AGN interpretation cannot be ruled out \citep{Perley16}. For these reasons, we do not include these two events in our sample of SLSNe-I.} and we add 15 new SLSNe-I. \cite{Inserra17} present a selection of \emph{Swift}-XRT observations of three SLSNe. The much longer temporal baseline and better sensitivity of the X-ray dataset presented here allow us to constrain the environments and the properties of central engines possibly powering the SLSN emission. 

We divide our sample into three groups: the ``gold sample'' (Table \ref{Tab:Gold}) contains 4 SLSNe-I with X-ray detections or well sampled optical bolometric light-curve and deep X-ray limits obtained with XMM or the CXO. The ``bronze sample'' contains 12 SLSNe-I with sparser optical data but with good \emph{Swift}-XRT X-ray coverage (Table \ref{Tab:Bronze}), while the ``iron sample'' comprises 10 SLSNe-I with very sparse optical and X-ray data (Table \ref{Tab:Iron}).  Given the peculiar nature of ASASSN-15lh \citep{Metzger15,Dong16,Leloudas16,vanPutten16,Chatzopoulos16,Dai15,Sukhbold16, Bersten16,Kozyreva16,Godoy-Rivera16,Margutti17}, this transient is not part of the sample of bona fide SLSNe-I analyzed here.  However, we discuss and compare the X-ray  properties of ASASSN-15lh in the context of SLSNe-I in Sec. \ref{Sec:IC}, \ref{Sec:GRB} and \ref{Sec:magnetar}.

%---------------------Swift-XRT/ CXO/ XMM
%The X-ray observations presented in this paper have been obtained with dedicated programs on \emph{Swift}-XRT, XMM and the CXO, as well as with serendipitous pointings of these spacecraft at the locations of SLSNe-I. 
\emph{Swift}-XRT data have been analyzed using  HEASOFT (v6.18) and corresponding calibration files, following standard procedures (see \citealt{Margutti13} for details). For each SLSN-I we provide stacked flux limits (for visualization purposes only), and flux limits derived from individual observations (Table \ref{Tab:master} in Appendix \ref{App}).  CXO data have been analyzed with the CIAO software package (v4.9) and corresponding calibration files. Standard ACIS data filtering has been applied. XMM data have been analyzed with SAS (v15.0). For the non-detections, we perform a flux calibration adopting a  power-law spectral model with index $\Gamma=2$ corrected for the Galactic neutral-hydrogen absorption along the line of sight (Tables \ref{Tab:Gold}, \ref{Tab:Bronze}, \ref{Tab:Iron}), as inferred from \cite{Kalberla05}. The details on X-ray observations of specific SLSNe-I are provided in Sec. \ref{SubSec:Gold} for the gold sample, and in Appendix \ref{App} for all the other SLSNe-I. Data tables can also be found in Appendix \ref{App}. Figure \ref{Fig:ddd} shows the complete sample of X-ray observations of SLSNe-I.

%%%%%%%%%%%%%%%%%%%%%%%%%%%%%%%%%%%%%%%%%%%
\subsection{Gold Sample}
\label{SubSec:Gold}

There are four objects in the gold sample:  SCP06F6, PTF12dam, PS1-14bj, and SN\,2015bn (Table \ref{Tab:Gold}).  In this section we describe the discovery and properties of each object in the gold sample.

%----------------------------------------
\subsubsection{SCP06F6}
X-ray emission at the location of the type-I SLSN SCP06F6 was first reported by \cite{Gansicke09} from XMM observations obtained 162 days after the initial detection of SCP06F6 (PI Shartel, ID 0410580301).  \cite{Levan13} derive an X-ray flux $F_x\sim10^{-13}\,\rm{erg\,s^{-1}cm^{-2}}$ (0.3-10 keV) on 2006 August 2 (MJD 53949, $\sim80$ days rest-frame since explosion). Follow-up observations with the CXO obtained on  2006 November 4 (MJD 54043, $\sim126$ days rest-frame since explosion, PI Murray, ID 7010) led to a non-detection. The corresponding flux limit is $F_x<1.4\times10^{-14}\,\rm{erg\,s^{-1}cm^{-2}}$ \citep{Levan13}. We adopt these values here and refer to \cite{Levan13} for further details. We note there that  the correct redshift for this event is $z=1.189$ \citep{Quimby11}.
%----------------------------------------
\subsubsection{X-ray emission at the location of PTF12dam}
\begin{figure}
\center
\includegraphics[scale=0.45]{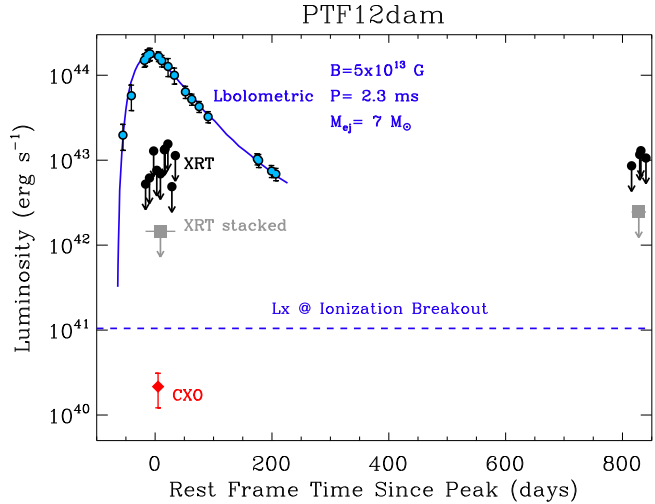}
\caption{Deep CXO observations (red diamond) obtained around the time of optical peak reveal the presence of soft X-ray emission at the location of PTF12dam with luminosity $L_x\sim2\times 10^{40}\,\rm{erg\,s^{-1}}$. Black filled circles: X-ray luminosity limits from \emph{Swift}-XRT. Grey filled squares: stacked limits from \emph{Swift}-XRT observations. Blue filled circles: bolometric optical emission as computed by \cite{Nicholl13}. Blue solid line: best-fitting magnetar model from \cite{Nicholl13} with parameters reported in Table \ref{Tab:Magnetar}. Horizontal blue dashed line: X-ray luminosity at the time of ionization breakout according to the equations in Sec. \ref{Sec:magnetar} for the best fitting magnetar parameters (Table \ref{Tab:Magnetar}). The expected time of ionization breakout is $t_{ion} = 4.7$ yr, see Table \ref{Tab:Magnetar}.}
\label{Fig:PTF12dam}
\end{figure}
PTF12dam (\citealt{Nicholl13,Chen15,Inserra17,Vreeswijk17}) belongs to the small subset of SLSNe-I with slowly evolving optical light-curves. At the time of writing, this group includes SN\,2007bi, PTF12dam, iPTF13ehe, SN2015bn, PS1-14bj and LSQ14an.  The slow evolution of these transients, and of  SN\,2007bi in particular, inspired a connection with pair instability explosions \citep{Gal-Yam09} that was later debated by \cite{Nicholl13}. X-ray observations of the SLSN-I PTF12dam  have been obtained with \emph{Swift}-XRT and the CXO.  \emph{Swift}-XRT observations span the time range $\sim43-900$ days rest-frame since explosion and revealed no detection down to a flux limit  $F_x\sim5\times10^{-14}\,\rm{erg\,s^{-1}cm^{-2}}$ (Fig. \ref{Fig:PTF12dam}, Table \ref{Tab:master}). 

A set of three deep CXO observations have been acquired between 2012 June 11 and June 19 ($\delta t\sim60-68$ days rest-frame since explosion; observations IDs 13772, 14444 and14446, PI Pooley). An X-ray source with a soft spectrum is clearly detected in the merged event file (total exposure of 99.9 ks) at the location of PTF12dam with significance of $4.8\sigma$ in the 0.5-2 keV energy range. The measured net count-rate is $(7.1\pm2.8)\times 10^{-5}\,\rm{c\,s^{-1}}$ (0.5-2 keV), which corresponds to an unabsorbed flux of $(7.3\pm2.9)\times 10^{-16}\,\rm{erg\,s^{-1}\,cm^{-2}}$ (0.3-10 keV) assuming a power-law spectrum with photon index $\Gamma=2$.  For a thermal bremsstrahlung spectrum with $T=0.24$ keV  (see below) the corresponding unabsorbed flux is $(8.9\pm3.5)\times 10^{-16}\,\rm{erg\,s^{-1}\,cm^{-2}}$ (0.3-10 keV), and $(3.5\pm1.4)\times 10^{-16}\,\rm{erg\,s^{-1}\,cm^{-2}}$ (0.5-2 keV).  In both cases, the inferred X-ray luminosity is $L_x\sim2\times 10^{40}\,\rm{erg\,s^{-1}}$  in the 0.3-10 keV (Fig. \ref{Fig:PTF12dam}).

PTF12dam exploded in a compact dwarf galaxy with fairly large star formation rate $SFR\sim5\,\rm{M_{\sun}\,yr^{-1}}$ \citep{Lunnan14,Chen15,Thone15,Leloudas15b,Perley16}\footnote{While \cite{Lunnan14}, \cite{Chen15}, \cite{Thone15} and \cite{Leloudas15b} measure $SFR\sim5\,\rm{M_{\sun}\,yr^{-1}}$, \cite{Perley16} report an even larger $SFR\sim10\,\rm{M_{\sun}\,yr^{-1}}$.}.  Following \cite{Mineo12}, the expected apparent diffuse X-ray emission associated with star formation  is 
$L_x/\rm{SFR}\approx 8.3\times10^{38}\,\rm{erg\,s^{-1}} (\rm{M_{\sun}\,yr^{-1}})^{-1}$, which translates into $L_x\approx 4.2 \times10^{39}\,\rm{erg\,s^{-1}} $  (0.5-2 keV) for $SFR\sim5\,\rm{M_{\sun}\,yr^{-1}}$. As a comparison, for a thermal bremsstrahlung spectrum with $T=0.24$ keV (average temperature of the unresolved X-ray component in galaxies, \citealt{Mineo12}), for PTF12dam we calculate $L_x\approx (1.0\pm0.4) \times10^{40}\,\rm{erg\,s^{-1}} $  (0.5-2 keV).  We therefore conclude that star formation in the host galaxy of PTF12dam is likely contributing to at least some of the X-ray luminosity that we detected at the location of the transient. In the following analysis sections we treat our measurements as upper limits to the X-ray emission from PTF12dam. These observations provide the deepest limits to the X-ray emission from a SLSN-I to date. Future observations will constrain the late-time behavior of the X-ray emission at the location of PTF12dam and will clarify its association to the optical transient. 

%----------------------------------------
\subsubsection{PS1-14bj}

\begin{figure}
\center
\includegraphics[scale=0.44]{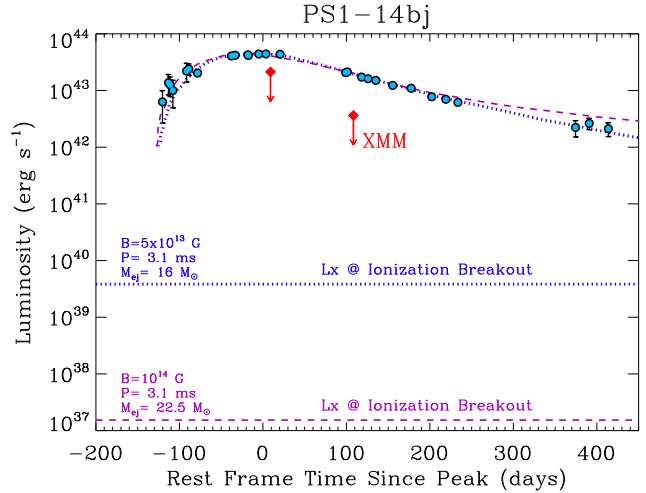}
\caption{Luminosity limits on the X-ray emission from PS1-14bj obtained with XMM (red diamonds). Blue filled circles: bolometric luminosity as computed by \cite{Lunnan16}. Blue dotted line and purple dashed line: magnetar models that adequately fit the observations as computed by \cite{Lunnan16} (see Table \ref{Tab:Magnetar}). Horizontal lines: X-ray luminosity at the time of ionization breakout according to the equations in Sec. \ref{Sec:magnetar} for the two magnetar models. For these models, the expected time of ionization breakout is $t_{ion} \geq 25$ yr, see Table \ref{Tab:Magnetar}.}
\label{Fig:14bj}
\end{figure}

Two epochs of deep X-ray observations of the type-I SLSN PS1-14bj \citep{Lunnan16} have been obtained with XMM (PI Margutti, IDs 0743110301, 0743110701) on 2014 June 9 ($\delta t\sim135$ days rest-frame since explosion, exposure time of 47.6 ks), and 2014 Nov 7 ($\delta t\sim235$ days rest-frame since explosion, exposure of 36.0 ks).  The net exposure times after removing  data with high background contamination are 3.6 ks and 29.8 ks, respectively (EPIC-pn data).  We do not find evidence for significant X-ray emission at the location of PS1-14bj in either observation and derive a $3\sigma$ 0.3-10 keV count-rate upper limit of $9.4\times 10^{-3}\,\rm{c\,s^{-1}}$ ($1.5\times 10^{-3}\,\rm{c\,s^{-1}}$)  for the first (second) epoch, which translates into an unabsorbed flux of $1.9\times 10^{-14}\,\rm{erg\,s^{-1}\,cm^{-2}}$  ($3.3\times 10^{-15}\,\rm{erg\,s^{-1}\,cm^{-2}}$).    The corresponding luminosity limits are shown in  Fig. \ref{Fig:14bj} and reported in Table \ref{Tab:master}. 
%----------------------------------------
\subsubsection{SN2015bn}
\begin{figure}
\includegraphics[scale=0.45]{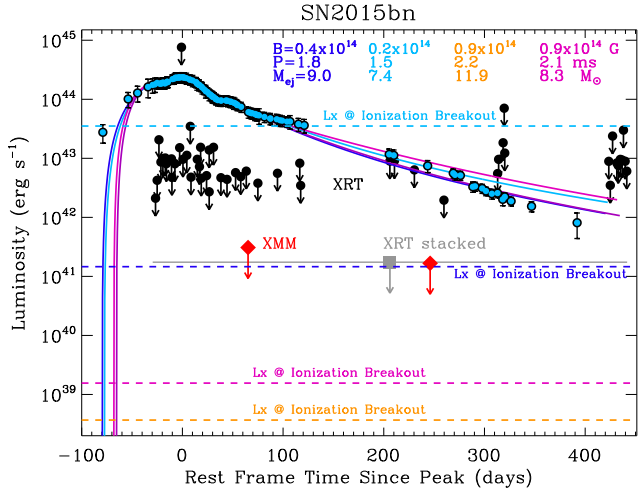}
\caption{Results from our joint \emph{Swift}-XRT and XMM X-ray campaign (black filled circles, grey squares and red diamonds) in the context of the optical bolometric luminosity of SN2015bn and the best-fitting magnetar models as derived by \cite{Nicholl16} (see Table \ref{Tab:Magnetar}). 
Horizontal dashed lines: expected X-ray luminosity at the time of the ionization breakout, which is $t_{ion}=5$ yrs (for  $B=0.4\times 10^{14}$ G, $P=1.8$ ms, $M_{\rm{ej}}=9\,\rm{M_{\sun}}$),  $t_{ion}=0.6$ yr (for  $B=0.2\times 10^{14}$ G, $P=1.5$ ms, $M_{\rm{ej}}=7.4\,\rm{M_{\sun}}$),   $t_{ion}=45$ yr (for  $B=0.9\times 10^{14}$ G, $P=2.2$ ms, $M_{\rm{ej}}=11.9\,\rm{M_{\sun}}$), $t_{ion}=22$ yr (for  $B=0.9\times 10^{14}$ G, $P=2.1$ ms, $M_{\rm{ej}}=8.3\,\rm{M_{\sun}}$) as reported in Table \ref{Tab:Magnetar}. The models with the shortest spin periods are disfavored by our X-ray limits. This figure clearly shows how magnetar models associated with very similar bolometric optical light-curves do predict instead very different X-ray luminosities at ionization breakout. The X-ray luminosity at the time of breakout is  a very sensitive probe of  the properties of central engines in SLSNe.}
\label{Fig:PS15aeLbol}
\end{figure}

X-ray observations of the SLSN-I 2015bn \citep{Nicholl16,Nicholl16b} have been obtained with \emph{Swift}-XRT and XMM (PI Margutti, IDs 0770380201, 0770380401). A first set of observations have been presented in \cite{Nicholl16b}, while \cite{Inserra17} include in their analysis five \emph{Swift}-XRT pointings. Here we present the complete data set. \emph{Swift}-XRT started observing SN2015bn on 2015 February 19 until 2016 July 23 covering the time period $\sim44-522$ days since explosion rest-frame. No statistically significant X-ray emission is blindly detected at the location of the transient (Fig. \ref{Fig:PS15aeLbol}).\footnote{We note the presence of marginally significant (2$\sigma$ c.l.) soft X-ray emission (i.e. $<0.3$ keV) with $L_x\sim5\times 10^{42}\,\rm{erg\,s^{-1}}$ found in a targeted search of data acquired on 2015 Feb 22 (i.e. $\sim55$ days since explosion, rest frame). However, emission with this flux is ruled out by \emph{Swift}-XRT observations obtained 24 hrs before, and is not detected in \emph{Swift}-XRT data with similar exposure time collected in the days afterwards. Furthermore, we find no evidence for X-ray emission when we filter the event file in the standard 0.3-10 keV energy range, which is where the \emph{Swift}-XRT is properly calibrated.  We conclude that the association of the targeted detection with real X-ray emission from SLSN-I 2015bn/PS15ae is highly questionable and therefore proceed with the conclusion of no statistically significant X-ray emission at the location of the transient.} 
 
Two epochs of XMM observations have been obtained on 2015 June 1 ($\delta t\sim145$ days rest frame since explosion) and 2015 Dec 18 ($\delta t\sim325$ days rest frame since explosion) with exposure times of 28.0 ks  and 25.1 ks, respectively (EPIC-pn data). After excluding time intervals heavily affected by proton flaring, the net exposure times are 7.3 ks and 18.8 ks. No X-ray source is detected at the location of the SLSN-I 2015bn.  We derive a $3\sigma$ 0.3-10 keV count-rate upper limit of $4.2\times 10^{-3}\,\rm{c\,s^{-1}}$ ($2.3\times 10^{-3}\,\rm{c\,s^{-1}}$)  for the first (second) epoch, which translates into an unabsorbed flux of $9.8\times 10^{-15}\,\rm{erg\,s\,cm^{-2}}$  ($5.3\times 10^{-15}\,\rm{erg\,s\,cm^{-2}}$). The results from  our X-ray campaign are listed in Table \ref{Tab:master} and displayed in Fig. \ref{Fig:PS15aeLbol}.

%%%%%%%%%%%%%%%%%%%%%%%%%%%%%%%%%%%%%%%%%%%
\section{Search for superluminous X-ray emission in SLSNe-I}
\label{Sec:SLXray}

\begin{figure}
\center
\includegraphics[scale=0.56]{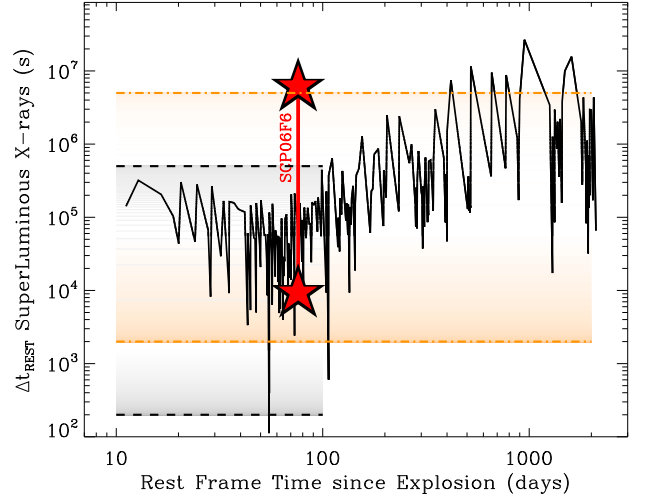}
\caption{Constraints on the duration of undetected superluminous X-ray emission  with $L_x\sim10^{45}\,\rm{erg\,s^{-1}}$ in SLSNe-I. Black  line: upper limit to the duration $\Delta t_{REST}$ of the (undetected) superluminous X-ray emission  from the analysis of the entire sample of SLSNe-I as a function of time since explosion. For SCP06F6, the range of allowed durations is between $\Delta t_{REST}\sim10^4$ s (exposure time of the XMM observation that provided a detection of X-ray emission) and the time of first detection $\Delta t_{REST}\sim70$ days (red stars). Horizontal black dashed lines: range of 3$\sigma$ c.l. allowed $\Delta t_{REST}$ based on the statistical analysis of observations of SLSNe-I in the first 100 days since explosion. Orange dot-dashed lines:  3$\sigma$ c.l. for $t<2000$ days since explosion. }
\label{Fig:duration}
\end{figure}

In this section we derive constraints on the possible presence of superluminous X-ray emission in SLSNe-I that was not detected because of the discontinuous observational coverage.  No assumption is made about the physical nature of the emission.  SLSNe-I are treated here as different realizations of the same stochastic process (which is the underlying assumption behind any sample analysis).

The hypothesis we test is that superluminous X-ray emission of the kind  detected at the location of SCP06F6 (i.e. $L_x\sim10^{45}\,\rm{erg\,s^{-1}}$) is ubiquitous in SLSNe-I. Our sample of observations comprises 253 spacecraft pointings, for a total observing time of $\sim 30$ days at $t<2000$ days (rest frame). Out of 253 trials, observations only led to one success (i.e. in the case of SCP06F6).  By using simple binomial probability arguments, we constrain the maximum and minimum $\Delta t_{REST}$ that would be statistically consistent at the 3$\sigma$ c.l. with 1 success out of $N$ trials, where $N\equiv N(t)$ and $N=253$ for $t=2000$ days. For $t<100$ days we find $200\,\rm{s}\leq\Delta t_{REST}\leq5\times 10^{5}\,\rm{s}$, while for $t<2000$ days we find $2000\,\rm{s}\leq\Delta t_{REST}\leq5\times 10^{6}\,\rm{s}$ (Fig. \ref{Fig:duration}, horizontal lines). In Figure \ref{Fig:duration}, we also show   with a thick black line the maximum duration of the superluminous X-ray emission that would be consistent with the lack of detections in our sample of SLSNe-I as a function of time since explosion. 

We conclude that superluminous X-ray emission is not a common trait of SLSNe-I. \emph{If} present, the superluminous X-ray emission requires peculiar physical circumstances to manifest and its duration is $\leq\,2$ months  at $t<2000$ days and $\leq$ few days at $t<100$ days.
%%%%%%%%%%%%%%%%%%%%%%%%%%%%%%%%%%%%%%%%%%%
\section{Constraints on SLSNe-I environments}
\label{Sec:IC}

\begin{figure*}
\center
\includegraphics[scale=0.5]{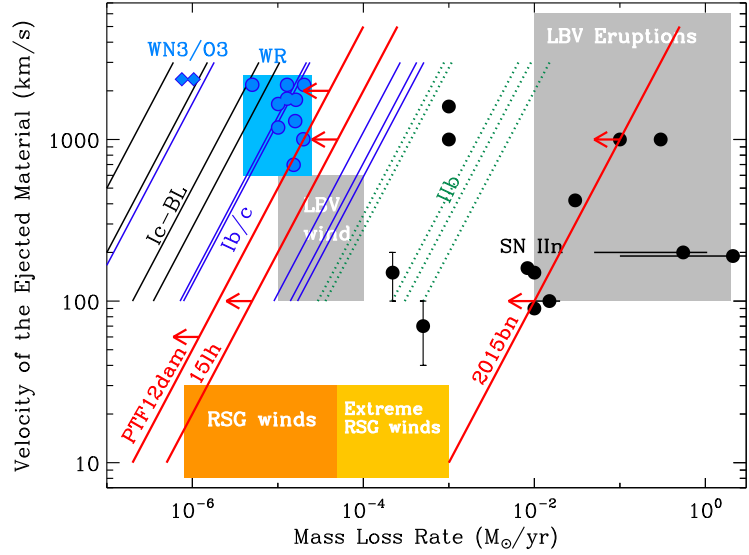}
\caption{Velocity of the ejected material during mass loss vs. pre-explosion mass-loss rate for H-poor core-collapse SNe (diagonal lines) and type-IIn SNe (black dots). H-poor SNe are represented with diagonal lines since radio and X-ray observations constrain the density $\rho_{CSM}$ which is $\propto \dot M/v_w$. 
Black, blue and dotted green lines mark the sample of type Ic-BL, Ib/c and IIb SNe from \citealt{Drout16}. SLSNe-I and the transient ASASSN-15lh are in red. 
For SLSNe-I we conservatively plot the constraints for $E_k=10^{51}\,\rm{erg}$, which is a \emph{lower limit} to the total kinetic energy of the blastwave (even in the case of a magnetar central engine). The properties of galactic WR stars are from \cite{Crowther07}, while WN3/O3 stars are from \cite{Massey15}. Locations of red supergiants environments (RSG) are from \cite{deJager88}, \cite{Marshall04} and \cite{vanLoon05}. Typical locations of Luminous Blue Variable (LBV) winds and eruptions are from \cite{Smith14} and \cite{Smith06}. The densest environments that characterize LBV eruptions and type-IIn SNe are not consistent with our deepest SLSNe-I limits. 
Our tightest constraints on PTF12dam rule out RSG winds and put PTF12dam in the same region of the parameter space as H-stripped SNe with broad spectral features (i.e. Ic-BL).}
\label{Fig:massloss1}
\end{figure*}

\begin{figure*}
\center
\includegraphics[scale=0.7]{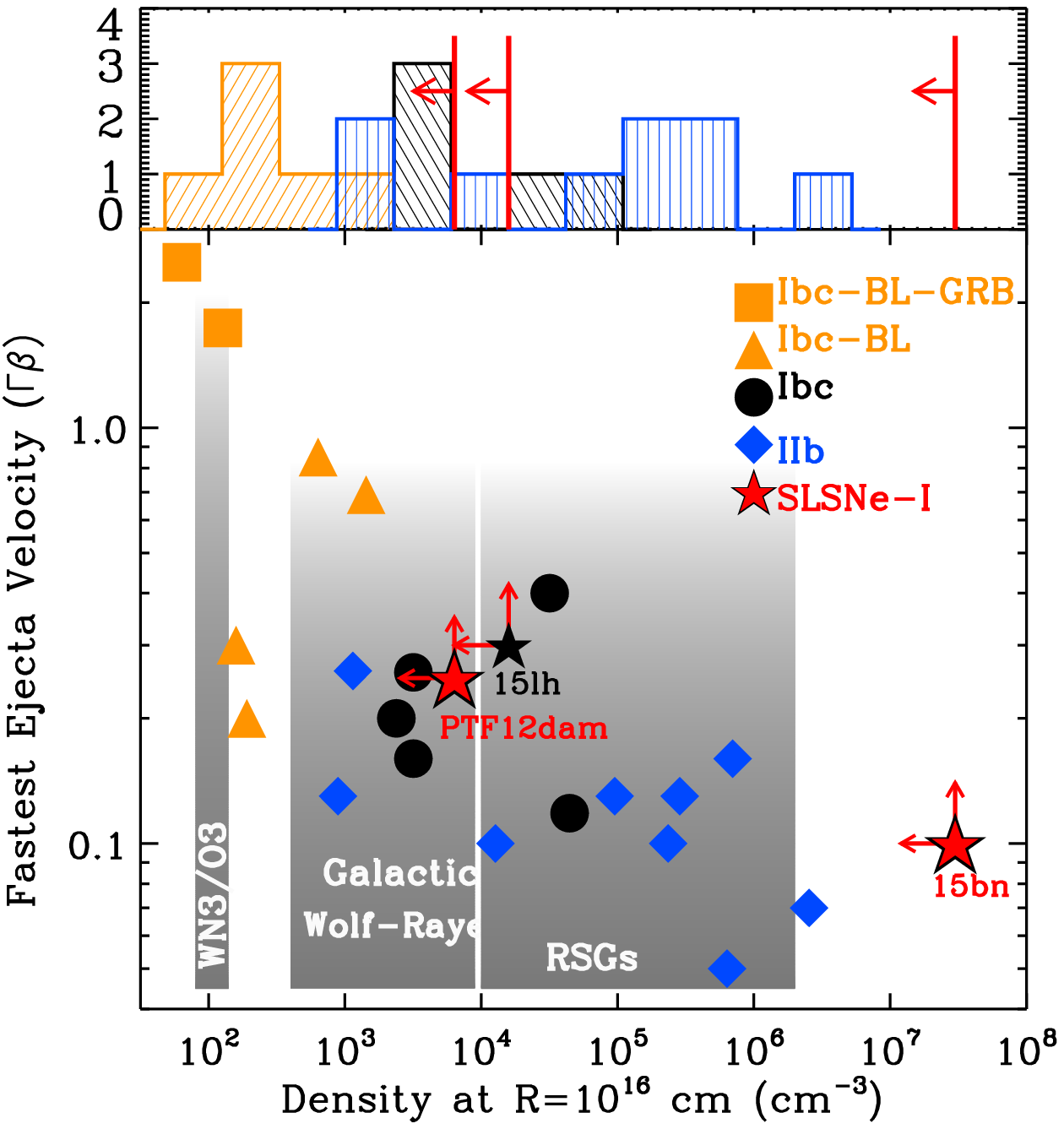}
\caption{Constraints on the fastest SN ejecta velocity and environmental density of SLSNe-I with the most sensitive X-ray limits (red stars) and the peculiar transient ASASSN-15lh (black star) in the context of core-collapse stellar explosions from H-stripped progenitors. Grey shaded regions: density in the environments of Red Supergiant Stars (RSGs), Wolf-Rayet stars (WRs) and the recently discovered new type of WR stars WN3/O3 (\citealt{deJager88,Marshall04,vanLoon05,Crowther07,Massey15}). Type IIb SNe (blue diamonds) explode in the densest environments, while SNe with broad spectroscopic features (orange squares and triangle) are associated with the lowest density media. For SLSNe-I we conservatively plot the constraints for $E_k=10^{51}\,\rm{erg}$, which is a \emph{lower limit} to the total kinetic energy of the blastwave (even in the case of a magnetar central engine). 
In the case of PTF12dam, our measurements rule out the dense environments associated with RSG winds and are consistent with the clean environments that characterize WRs, WN3/O3 and engine-driven SN explosions. References: \cite{vanDyk94,Fransson98,Berger02,Weiler02,Ryder04,Soderberg05,Chevalier06,Soderberg06b,Soderberg06c,Soderberg08,Roming09,Soderberg10,Soderberg10b,Krauss12,Milisavljevic13,Margutti14b,Kamble14,Corsi14,Chakraborti15,Drout16,Kamble16,Margutti17}.}
\label{Fig:massloss2}
\end{figure*}

Inverse Compton (IC) emission is a well known source of X-rays in young stellar explosions \citep{Bjornsson04,Chevalier06}. X-ray emission originates from the up-scattering of optical photons from the SN photosphere by a population of relativistic electrons accelerated at the shock front. While always present, IC is the dominant emission mechanism at early times ($t\leq$ optical peak) for SN propagating into low-density media. In the case of strong SN shock interaction with the medium, the dominant X-ray emission mechanism is instead bremsstrahlung \citep{Bjornsson04,Chevalier06} as it is indeed observed in type-IIn SNe and as recently confirmed by the first broad-band X-ray spectra of strongly interacting SNe (\citealt{Ofek14,Margutti17}). The analysis of the optical emission from SLSNe-I in the context of the interaction model  (e.g. \citealt{Nicholl14, Nicholl16}) suggests that \emph{if} SLSNe-I are powered by interaction then the shock breaks out around the time of optical maximum light and  the medium consists of a thick shell confined to small radii ($R\sim5\times10^{15}\,\rm{cm}$ for SN2015bn) surrounded by a lower density region. This conclusion is consistent with the lack of observed narrow lines in the optical spectra of SLSNe-I (in sharp contrast to ordinary and superluminous type-IIn SNe): the presence of an extended unshocked region of dense CSM would likely imprint low-velocity features which are \emph{not} observed in SLSNe-I (see also \citealt{Chevalier11}). %\footnote{The SLSN-IIn 2006gy is maybe the most extreme example of SLSN with strong signs of interaction. From the modeling of the optical emission \textcolor{red}{Smith 2006} inferred   }
The X-ray observations that we will use in this section have been obtained at the time of maximum light or later, which is after the shock has broken out from the thick shell of material \emph{if} a shell is there. In the following, we thus constrain the density around SLSNe-I under the conservative assumption that IC is the \emph{only} source of X-ray radiation.  Since we sample the time range $t>t_{peak}$, our  density limits apply to the region $R\gtrsim10^{16}\,\rm{cm}$. 

The X-ray emission from IC depends on (i) the density structure of the SN ejecta and of the circum-stellar medium (CSM), (ii) the details of the electron distribution responsible for the up-scattering, (iii) the explosion parameters (ejecta mass $M_{ej}$ and kinetic energy $E_k$), and (iv) the availability of seed optical photons ($L_{x,IC}\propto L_{bol}$, where $L_{bol}$ is the bolometric optical luminosity). We employ the formalism of \cite{Margutti12} modified to reflect the stellar structure of massive stars as in \cite{Margutti14b}. We further assume a wind-like medium with $\rho_{CSM}\propto R^{-2}$ as appropriate for massive stars, a power-law electron distribution $n_{e}(\gamma)=n_0 \gamma^{-p}$ with 
$p\sim3$ as indicated by radio observations of H-stripped core-collapse SNe \citep{Chevalier06} and a fraction of post-shock energy into relativistic electrons  $\epsilon_e=0.1$ (e.g.  \citealt{Chevalier06}). Since $L_{x,IC}\propto L_{bol}$, it is clear that the tightest constraints on $\rho_{CSM}$ will be derived from the most nearby SLSNe-I, which have very bright optical emission and deep X-ray limits  (i.e. they have the largest flux ratio $F_{opt}/F_x$ constrained by observations). To this end, we analyze below the SLSNe-I 2015bn and PTF12dam. We also provide constraints for the peculiar transient ASASSN-15lh.

For SN2015bn we follow \cite{Nicholl16b} and \cite{Nicholl16} and adopt a range of ejecta masses $M_{ej}=7-15\,\rm{M_{\sun}}$ (Table \ref{Tab:Magnetar}). With these parameters and the optical bolometric light-curve from \cite{Nicholl16} (Fig. \ref{Fig:PS15aeLbol}), our X-ray non detections constrain the pre-explosion mass-loss rate from the stellar progenitor of SN2015bn to $\dot M<10^{-2}\,\rm{M_{\sun}yr^{-1}}$ ($\dot M<10^{-1}\,\rm{M_{\sun}yr^{-1}}$) for  $E_k=10^{52}\,\rm{erg}$ ($E_k=10^{51}\,\rm{erg}$) and wind velocity $v_w=1000\,\rm{km\,s^{-1}}$, (Fig. \ref{Fig:massloss1},  \ref{Fig:massloss2}), which is $\dot M<10^{-4}\,\rm{M_{\sun}yr^{-1}}$  ($\dot M<10^{-3}\,\rm{M_{\sun}yr^{-1}}$) for wind velocity $v_w=10\,\rm{km\,s^{-1}}$.  In this context, the analysis of the radio observations of SN2015bn indicates $\dot M<10^{-2}\,\rm{M_{\sun}yr^{-1}}$ for $v_w=10\,\rm{km\,s^{-1}}$ at $R>10^{15}\,\rm{cm}$, while  $\dot M\sim10^{-2}\,\rm{M_{\sun}yr^{-1}}$ would be needed to explain the late-time optical  light  curve of the transient through continued ejecta-CSM  interaction  \citep{Nicholl16}. The X-ray analysis thus argues against the presence of an extended CSM region if $E_k>10^{51}\,\rm{erg}$ (as it is likely the case)  and suggests that another source of energy is powering the light-curve after peak. This result is consistent with the conclusions by \cite{Nicholl16b}: based on the spectroscopic similarity of SN2015bn with the GRB SN1998bw in the nebular phase,  \cite{Nicholl16b} concluded that a central engine is driving the explosion.

For PTF12dam we use $M_{ej}=7\,\rm{M_{\sun}}$ as inferred by \cite{Nicholl13} from the modeling of the optical bolometric emission (Table \ref{Tab:Magnetar}). We detect an X-ray source at the location of PTF12dam with $L_x\sim2\times 10^{40}\,\rm{erg\,s^{-1}}$. We treat this value as an upper limit to the X-ray luminosity from the transient to account for possible contamination from the host galaxy.
For these values of the explosion parameters and the measured $L_x$, the inferred mass-loss rate is $\dot M<2\times10^{-5}\,\rm{M_{\sun}yr^{-1}}$ ($\dot M<4\times10^{-6}\,\rm{M_{\sun}yr^{-1}}$) for $E_k=10^{51}\,\rm{erg}$ ($E_k=10^{52}\,\rm{erg}$) and $v_w=1000\,\rm{km\,s^{-1}}$ (Fig. \ref{Fig:massloss1},  \ref{Fig:massloss2}).  These are the tightest constraints to the pre-explosion mass-loss history of SLSNe-I progenitors.

\emph{If} the peculiar transient ASASSN-15lh  is associated with an $E_k=10^{52}\,\rm{erg}$ explosion with ejecta mass $M_{ej}=5-10\,\rm{M_{\sun}}$ (\citealt{Metzger15,Chatzopoulos16,Dong16,Kozyreva16,Sukhbold16,Bersten16,Dai+16}), the X-ray observations from \cite{Margutti16}, imply $\dot M<5\times10^{-6}\,\rm{M_{\sun}yr^{-1}}$  ($\dot M<5\times10^{-5}\,\rm{M_{\sun}yr^{-1}}$) for $v_w=1000\,\rm{km\,s^{-1}}$  for a thermal (non thermal) X-ray spectrum. These values are typical of mass-loss rates from H-stripped compact massive stars (Fig. \ref{Fig:massloss1},  \ref{Fig:massloss2}).

We end by noting that the calculations of the expected X-ray luminosity from the strong interaction scenario applied to SLSNe-I reported in \cite{Inserra17} provide \emph{upper limits} to the X-ray luminosity from the interaction scenario rather than expected $L_x$ as it is assumed that as much as 10\% of the  total kinetic energy is converted into X-ray photons in the 0.3-10 keV energy range that would leak out and reach the observer without being absorbed, thermalized and/or downgraded to lower energies due to scattering processes within the dense media. For this reason, the large $L_x\sim10^{43}\,\rm{erg/s}$ predicted with this method by \cite{Inserra17} cannot be directly compared to the observed X-ray limits and cannot be used to rule out the strong interaction scenario. As a matter of fact, the SLSN-IIn 2006gy is maybe the most extreme example of SLSN powered by interaction with a massive CSM shell with $M\sim 10\,\rm{M_{\sun}}$ and it was detected in the X-rays around optical maximum light with $L_x\sim10^{39}\,\rm{erg/s}$ which is clearly $\ll 10^{43}\rm{erg\,s^{-1}}$ \citep{Smith07b}.

From our analysis, we conclude that for  PTF12dam the inferred limits rule out the densest environments that characterize type-IIn SNe (Fig. \ref{Fig:massloss1}), indicating that the strong SN shock interaction with an \emph{extended} medium is unlikely to be the primary source of energy sustaining the very luminous display. Interestingly, a  low-density environment with $\dot M\sim 4\times 10^{-6}\,\rm{M_{\sun}yr^{-1}}$ was also inferred from radio and X-ray observations of the type-Ib SN\,2012au \citep{Kamble14}, which showed spectroscopic similarities with SLSN-I \citep{Milisavljevic13}. The tight constraints obtained for PTF12dam point to a clean environment, and argue against the dense CSM typical of extended progenitors like RSG stars.  This result suggests that  at least some SLSN-I progenitors are likely to be compact stars surrounded by a low density medium at the time of stellar death.
%%%%%%%%%%%%%%%%%%%%%%%%%%%%%%%%%%%%%%%%%%%
\section{Central engines in SLSNe-I}
\label{Sec:CE}
%------------------------------------------------------------------------------------------------------------------
\subsection{Constraints on on-axis and off-axis collimated and non-collimated relativistic outflows}
\label{Sec:GRB}
\begin{figure*}
\center
\includegraphics[scale=0.7]{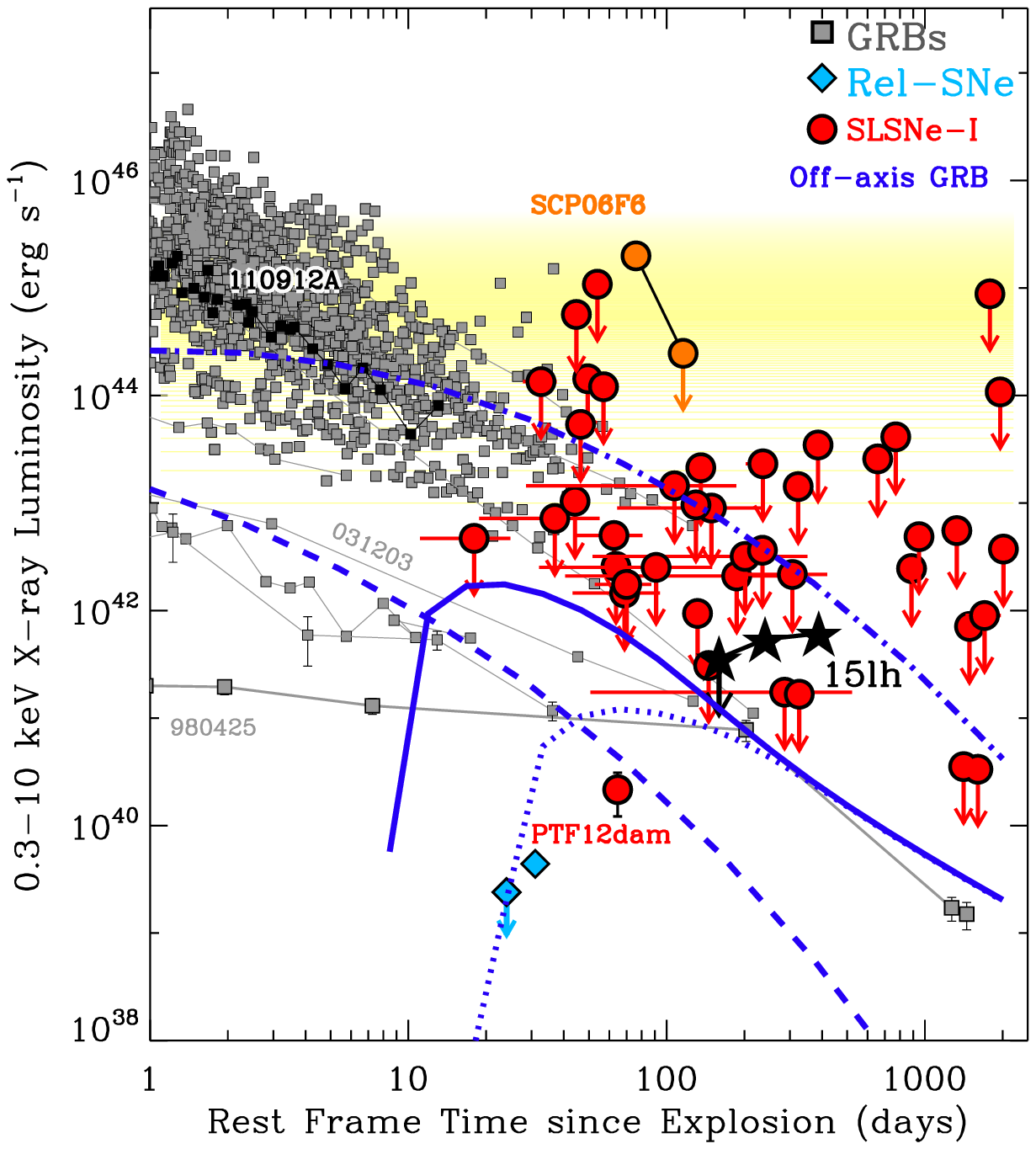}
\caption{X-ray emission from SLSNe-I (red circles) in the context of detected GRB X-ray afterglows (grey squares, \citealt{Margutti13}), relativistic SNe (blue diamonds, \citealt{Soderberg10,Margutti14b}) and representative off-axis afterglow models (blue lines) from collimated outflows with $\theta_{\rm{jet}}=5^{\circ}$, $\epsilon_{\rm{e}}=0.1$,  $\epsilon_{\rm{B}}=0.01$, $E_{\rm{k}}=4\times 10^{52}\,\rm{erg}$, $\dot M=10^{-3}\,\rm{M_{\sun}yr^{-1}}$, $\theta_{\rm{obs}}=30^{\circ}$ (thick line) and $\theta_{\rm{obs}}=45^{\circ}$ (dotted line). We also show models for $\dot M=10^{-7}\,\rm{M_{\sun}yr^{-1}}$, $\theta_{\rm{obs}}=2\theta_{\rm{jet}}$, $E_{\rm{k}}=4\times 10^{52}\,\rm{erg}$ (dot-dashed line) and $E_{\rm{k}}=4\times 10^{50}\,\rm{erg}$ (dashed line). The fast-fading X-ray emission at the location of SCP06F6 is shown with orange circles \citep{Levan13}. Black stars: steady X-ray emission at the location of ASASSN-15lh (\citealt{Margutti16}). Black squares: X-ray afterglow of GRB\,111209A, associated with the SLSN 2011kl \citep{Greiner15}.  In this plot we show the \emph{Swift}-XRT limits from the stacked analysis for displaying purposes. The analysis and results are based on the time-resolved observations. Notably, our deepest limits rule out non-collimated outflows from the weakest GRB explosions, like GRB\,980425 (\citealt{Pian00,Kouveliotou04}).}
\label{Fig:GRBoff}
\end{figure*}
The search for off-axis and on-axis relativistic GRB-like jets in  SLSNe-I  is motivated by two recent observational findings: (i) the association of SN2011kl with GRB111209A. SN2011kl bridges the luminosity gap between GRB-SNe and SLSNe-I, and shows spectroscopic similarities to SLSNe-I \citep{Greiner15}.  (ii) Nebular spectroscopy of the SLSN-I 2015bn revealed close similarities to the engine-driven SN1998bw, associated with GRB\,980425, suggesting that the core of engine-driven SNe and SLSNe-I share some key physical properties and structure \citep{Nicholl16b}.

Early-time X-ray observations of SLSNe-I acquired at $t\lesssim40$ days generally rule out on-axis collimated ultra-relativistic outflows of the type associated with energetic long GRBs (Fig. \ref{Fig:GRBoff}, cloud of filled grey squares).  We constrain the presence of off-axis relativistic outflows by generating a grid of off-axis GRB X-ray afterglows with the broadband afterglow numerical code Boxfit v2 \citep{vanEerten12}.\footnote{http://cosmo.nyu.edu/afterglowlibrary/boxfit2011.html} The observed X-ray emission depends on the kinetic energy $E_k$ of the outflow, the density of the medium $\rho_{CSM}$ (we explore both an ISM-like medium $n_{CSM}=const$ and a wind-like medium with $\rho_{CSM}\propto R^{-2}$), the microphysical shock parameters $\epsilon_B$ and $\epsilon_e$ (postshock energy fraction in magnetic field and electrons, respectively), the jet opening angle $\theta_j$ and the angle of the jet with respect to the line of sight $\theta_{obs}$. We explore the predicted X-ray signatures of collimated $\theta_j=5^{\degree}$ outflows with $\epsilon_e=0.1$ and $\epsilon_B=0.01$ (as derived from first-principle simulations of relativistic shocks, e.g. \citealt{Sironi15}), isotropic kinetic energy in the range $E_{k,iso}=10^{52}-10^{55}\,\rm{erg}$, environment density in the range $n=10^{-3}-10\,\rm{cm^{-3}}$  (ISM) or mass-loss rate $\dot M=10^{-7}-10^{-3}\,\rm{M_{\sun}\,yr^{-1}}$ (wind) and observed angles $\theta_{obs}\leq 90^{\degree}$. These values are representative of the parameters inferred from accurate modeling of broad-band afterglows of GRBs. 

Based on these simulations and the X-ray observations from the entire sample of SLSNe-I, we find that relativistic collimated outflows with $E_k>10^{51}\,\rm{erg}$, $n>10^{-3}\,\rm{cm^{-3}}$ and  $\theta_{obs}<2\theta_j$ are ruled out. Powerful jets with $E_k>10^{52}\,\rm{erg}$ expanding in a thick medium with $n\geq 10\,\rm{cm^{-3}}$ and $\theta_{obs}\leq 30^{\degree}$ are also ruled out. For a wind environment, our observations are not consistent with jets with $E_k>10^{50.5}\,\rm{erg}$ expanding in a medium enriched with $\dot M\geq10^{-7}\,\rm{M_{\sun}\,yr^{-1}}$ and $\theta_{obs}<2\theta_j$. At higher kinetic energies  observations rule out jets with $E_k>10^{51.5}\,\rm{erg}$, $\dot M\geq10^{-4}\,\rm{M_{\sun}\,yr^{-1}}$ and $\theta_{obs}\leq 30^{\degree}$ or $E_k>10^{52}\,\rm{erg}$, $\dot M\geq10^{-3}\,\rm{M_{\sun}\,yr^{-1}}$ and $\theta_{obs}\leq 45^{\degree}$. We are not sensitive to jets viewed at $\theta_{obs}>30^{\degree}$ for the ISM, and $\theta_{obs}>45^{\degree}$ for the wind medium.

In the case of PTF12dam, observations argue against jets with $E_k>10^{51}\,\rm{erg}$ propagating into a medium with $n>10^{-3}\,\rm{cm^{-3}}$ or  $\dot M>10^{-7}\,\rm{M_{\sun}\,yr^{-1}}$ and $\theta_{obs}<2\theta_j$. The portion of the parameter space associated with $E_k=10^{50.5}\,\rm{erg}$, $\dot M\sim10^{-6-7}\,\rm{M_{\sun}\,yr^{-1}}$  and $\theta_{obs}<2\theta_j$ is also ruled out. Dense environments with $n>10\,\rm{cm^{-3}}$ or  $\dot M>10^{-4}\,\rm{M_{\sun}\,yr^{-1}}$ would also produce X-ray emission in excess to what we observed for outflows with $E_k>10^{51}\,\rm{erg}$ viewed at  $\theta_{ob s}<30^{\degree}$.  

For the SLSN-I 2015bn observations rule out systems with $E_k>10^{52}\,\rm{erg}$, $n>10^{-3}\,\rm{cm^{-3}}$ or $\dot M>10^{-7}\,\rm{M_{\sun}\,yr^{-1}}$ for $\theta_{obs}<2\theta_j$. Even the most energetic outflows in our simulations with $E_k>10^{52}\,\rm{erg}$ would fall below our detection threshold for $\theta_{obs}>30^{\degree}$ and the range of densities considered. These observations complement the results  from deep radio non-detections of SN2015bn \citep{Nicholl16}, which argue against powerful on-axis or off-axis jets with $E_k=2\times10^{51}\,\rm{erg}$ propagating into an ISM-like medium with density $n=1\,\rm{cm^{-3}}$.\footnote{Note that \cite{Nicholl16} assume an ISM-like medium and larger $\epsilon_B=0.1$ and $\theta_j=10\degree$.}

Finally, we consider the observable X-ray signatures of \emph{non-collimated} mildly-relativistic outflows. X-ray observations of the majority of SLSNe-I in our sample are not sensitive to the faint X-ray emission of mildly-relativistic non-collimated outflows typical of low-energy GRBs like 980425, 031203, 060218 and 100316D (e.g. \citealt{Pian00,Kouveliotou04,Watson04,Soderberg06c,Margutti13b}). However, our deepest X-ray limits obtained with the CXO and XMM are sensitive enough to probe the parameter space populated by the weakest GRB-SNe.  For the SLSN-I 2015bn, our XMM observations probe and rule out luminosities $L_x>2\times 10^{41}\,\rm{erg\,s^{-1}}$ at $t\sim100-300$ days, which are comparable to the detected X-ray emission of GRBs 980425 and 031203 at a similar epoch, $L_x\sim 10^{41}\,\rm{erg\,s^{-1}}$ (Fig. \ref{Fig:GRBoff}).  
Remarkably, in the case of PTF12dam, CXO observations acquired at the time of optical peak rule out even the faintest non-collimated X-ray emission ever detected from a low-energy GRB (Fig. \ref{Fig:GRBoff}), indicating that, \emph{if} PTF12dam is an engine-driven stellar explosion, the jet never successfully broke out from the stellar envelope, in close analogy to the picture recently suggested for the relativistic SNe 2009bb and 2012ap (\citealt{Margutti14b} and references therein). 

To conclude, the analysis of our deep X-ray limits in the context of GRB afterglows simulations and the recent finding of similarity in the nebular emission from SN2015bn with engine-driven SNe, suggest that either SLSNe-I are powered by very energetic collimated GRB-like outflows that were pointing far away from our line of sight ($\theta_{ob s}>30^{\degree}$), or that SLSNe-I harbor failed jets that do not successfully break through the stellar envelope and are associated with weak X-ray emission. Late-time radio observations of SN2015bn \citep{Nicholl16b} argue against the off-axis relativistic jet scenario. However, the association of SN2011kl with GRB111209A clearly implies that at least some SLSNe-I harbor relativistic jets. We therefore propose that, in strict analogy to H-stripped core-collapse SNe of ordinary luminosity (e.g. \citealt{Xu08, Mazzali08, Lazzati12,Margutti14b}), SLSNe-I are also characterized by a continuum of jets strengths and life-time of the central engine.

%------------------------------------------------------------------------------------------------------------------
\subsection{Constraints on magnetar central engines: the Ionization Break out}
\label{Sec:magnetar}

\begin{figure*}
\center
\includegraphics[scale=0.7]{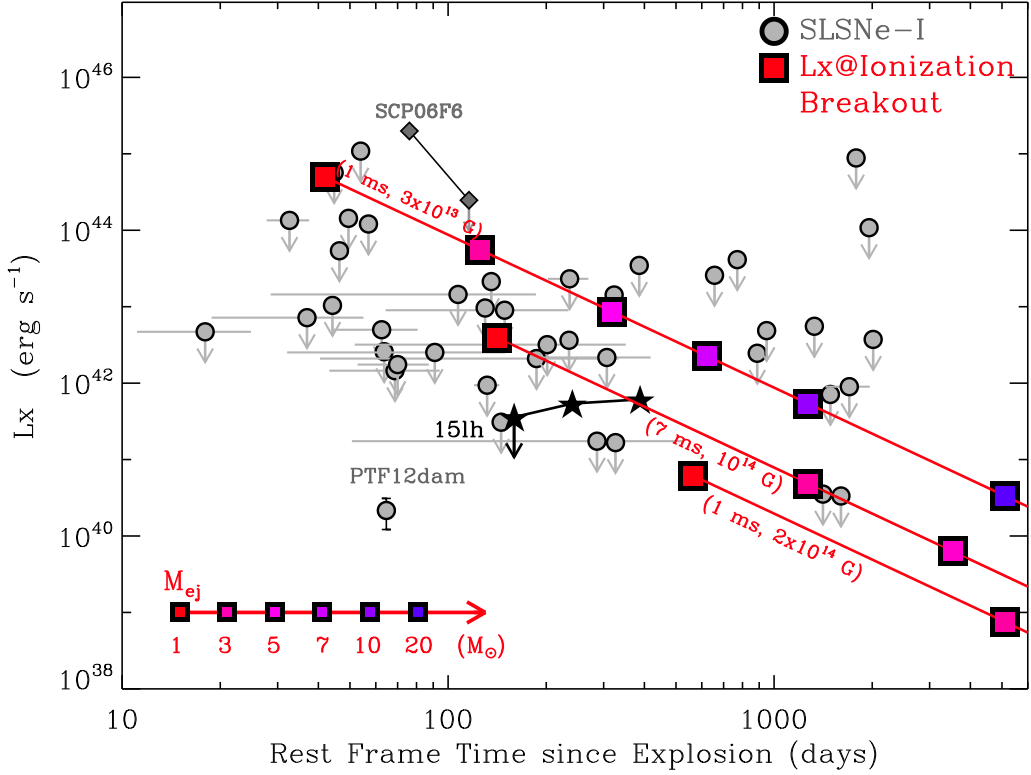}
\caption{X-ray luminosity at the time of ionization break out (thick red line) for a variety of representative magnetar parameters, $P=1$ ms, $B=3\times 10^{13}$ G (upper line), $P=7$ ms, $B=10^{14}$ G (middle line), $P=1$ ms, $B=2\times10^{14}$ G (lower line), and for a range of ejecta mass values $M_{\rm{ej}}$ between 1 and 20 $M_{\sun}$. These calculations assume an Oxygen dominated ejecta composition ($Z_8=1$), $v=10^9$ cm/s, $T=10^5$ K, $X_A=0.1$. Our limits (grey dots) rule out the fastest spinning magnetars with $P\le 7$ ms, $B\le 10^{14}$ G and small ejecta masses $M_{\rm{ej}}\le5\,\rm{M_{\sun}}$. However, we are not sensitive to magnetars with $B_{14}\ge2\times 10^{14}$ G.  Grey diamonds: X-ray emission at the location of SCP06F6 \citep{Levan13}. Black stars: X-ray emission at the location of ASASSN-15lh \citep{Margutti16}.}
\label{Fig:magnetar}
\end{figure*}

We compute the ionization breakout time and the X-ray luminosity at breakout following \cite{Metzger14} and \cite{MetzgerPiro14}. We consider a central engine with an UV/X-ray luminosity $L$ that releases an energy $L \times t$ in ionizing radiation on a timescale $t$. 
The radiation ionizes its way through the ejecta on a timescale 
\begin{eqnarray}
t_{\rm ion} \approx  \left\{
\begin{array}{lr}
120\,{\rm d}\,M_3^{3/4} v_9^{-5/4}T_{5}^{-0.2}\left(\frac{X_{A}}{0.1}\right)^{1/4}\left(\frac{L t}{10^{52}{\rm ergs}}\right)^{-1/4}Z_8^{3/4}
, 
(\eta_{\rm thr} \ll 1) \\
110\,{\rm d}\,M_3 v_9^{-3/2}T_{5}^{-0.4}\left(\frac{X_{A}}{0.1}\right)^{1/2}\left(\frac{L t}{10^{52}{\rm ergs}}\right)^{-1/2}Z_8^{3/2},       
(\eta_{\rm thr} \gg 1), \\
\end{array}
\right.
\label{eq:tbo}
\end{eqnarray}
where $M_3 \equiv M_{\rm ej}/(3M_{\odot})$, $T_{5} = T/10^{5}$ K is the temperature of electrons in the recombination layer, $v_9\equiv v/10^9\,\rm{cm\,s^{-1}}$, $X_Z$ is the mass fraction $X_Z$ of elements with atomic number $Z = 8Z_8$ in the ejecta and
\begin{eqnarray}
\eta_{\rm thr} \approx 0.7\left(\frac{L t}{10^{52}{\rm erg}}\right)^{-1}M_{3}v_9^{-1}\left(\frac{X_{A}}{0.1}\right)T_{5}^{-0.8}Z_8^{3}
\label{eq:etaA}
\end{eqnarray}
is the ratio of absorptive and scattering opacity in the ejecta \citep{Metzger14}. 
The spin-down timescale $t_{\rm{sd}}$ of a magnetar central engine is given by
\begin{equation}
t_{\rm{sd}}\sim 4.7\rm{d}B_{14}^{-2}P^2_{\rm{ms}}
\end{equation}
and the spin-down luminosity is given by
\begin{equation}
L_{\rm{sd}}=5\times 10^{46}B_{14}^{2}P^{-4}_{\rm{ms}}\Big (1+\frac{t}{t_{\rm{sd}}} \Big )^{-2}\,\rm{erg\,s^{-1}}\approx 1.1\times 10^{48} B_{14}^{-2}t_{\rm{d}}^{-2} \,\rm{erg\,s^{-1}}
\end{equation}
for $t\gg t_{\rm{sd}}$,  $B_{14}\equiv B/10^{14}\,\rm{G}$, $P_{\rm{ms}}\equiv P/\rm{ms}$, $t_{\rm{d}}\equiv t/\rm{days}$ and we adopted the vacuum dipole spin-down convention employed by \cite{Kasen10}.
For $L=L_{\rm{sd}}$ and $t\gg t_{\rm{sd}}$ the ionization time scale of Eq. \ref{eq:tbo} can be written as follows:
\begin{eqnarray}
t_{\rm ion} \approx  \left\{
\begin{array}{lr}
280\,{\rm d}\,M_3 v_9^{-5/3}T_{5}^{-4/15}(\frac{X_A}{0.1})^{1/3} B_{14}^{2/3}Z_8, 
(\eta_{\rm thr} \ll 1) \\
1273\,{\rm d}\,M_3^{2} v_9^{-3} T_{5}^{-4/5} (\frac{X_A}{0.1}) B_{14}^2 Z_8^{3},       
(\eta_{\rm thr} \gg 1), \\
\end{array}
\right.
\label{eq:tbo2}
\end{eqnarray}
The X-ray luminosity at ionization breakout is $L_x\approx L_{\rm{sd}}(t_{\rm{ion}})/14$.  In the following we assume that Oxygen ($Z_8=1$) dominates the bound-free opacity in the $\sim$keV X-ray band, we use and electron temperature $T=10^5$ K, $X_A=0.1$, velocity of the order of $10^9$ cm/s and we compute the ionization time scale and the X-ray luminosity at break out and compare to our X-ray limits and measurements. 

For SLSNe-I with well constrained optical bolometric emission, we use the magnetar parameters $M_{ej}$, $P$ and $B$ that best fit the optical bolometric luminosity to estimate $t_{ion}$ and $L_x(t_{ion})$ (Table \ref{Tab:Magnetar}). From Table \ref{Tab:Magnetar} and Fig. \ref{Fig:PTF12dam}-\ref{Fig:PS15aeLbol} it is clear that most of the $L_x(t_{ion})$ are too faint to be detected and that $t_{ion}$ is usually much larger than the $\sim2000$ days that we cover with our observations. However, it is also clear that $t_{ion}$ and  $L_x(t_{ion})$ are very sensitive to the magnetar parameters and qualify as excellent probes of central engines in SLSNe-I.  
%Figure \ref{Fig:PS15aeLbol}  shows how magnetar parameters that provide very similar fits to the observed optical bolometric emission do instead predict very different X-ray luminosities at ionization breakout ($L_x(t_{ion})$ varies of a factor $\sim10^5$ among the models that best-fitting models)
Magnetar central engines that would produce very similar optical bolometric outputs that cannot be distinguished with current optical-UV photometry, are instead clearly differentiated in their $L_x(t_{ion})$ - $t_{ion}$ properties. 
As an example, for the best fitting magnetar parameters of SN2015bn in Fig. \ref{Fig:PS15aeLbol}, $L_x(t_{ion})$ spans $\sim5$ orders of magnitude and $t_{ion}$ ranges from $0.6-88$ yrs. For this SLSN-I, our deep X- ray limits obtained with XMM and the combined limit from \emph{Swift}-XRT favor models with $P > 2$ ms. The fastest spinning magnetar model with $P = 1.5$ ms and relatively small ejecta mass $M_{ej}=7.4\,\rm{M_{\sun}}$  from \cite{Nicholl16b} predicts $L_x(t_{ion})>10^{43}\,\rm{erg\,s^{-1}}$ at $t_{ion}\sim0.6$ yrs and it is therefore disfavored by our X-ray observations (Fig. \ref{Fig:PS15aeLbol}). %\textbf{Figure \ref{Fig:PS15aeLbol} shows that the bolometric light-curve at late times falls more steeply than expected from the magnetar models, as noted by \cite{Nicholl16b}, with an ``excess'' of luminosity $L\sim1.4\times 10^{42}\,\rm{erg\,s^{-1}}$ at $\sim350-400$ days rest-frame since peak. If this energy is leaking out in the X-rays, the unabsorbed $F_x\sim4\times 10^{-14}\,\rm{erg\,s^{-1}cm^{-2}}$ (0.3-10 keV). Collect XRT data after 2016 July 3, which is MJD=57572 }.

Figure \ref{Fig:magnetar} shows how the X-ray observations from our sample compare to the predictions from the magnetar ionization breakout model. We investigate a wide range of central engine parameters $P=1-7$ ms and $B_{14}=0.2-10$ G for ejecta masses $M_{ej}=1-20\,\rm{M_{\sun}}$.  Current X-ray observations are not sensitive to magnetars with $B_{14}\geq 2$ (Fig. \ref{Fig:magnetar}). For $B_{14}< 2$,  observations favor models with larger ejecta mass: for $B_{14}=0.2,0.5,1.0$ G the allowed parameter space is $M_{ej}>20,7,3\,\rm{M_{\sun}}$. X-ray observations indicate that \emph{if} a magnetar central engine powers the emission from SLSNe-I then it has to be either associated with a large magnetic field or with a large explosion ejecta mass. These results are independent from (but in agreement with) the values inferred from the modeling of the optical emission from SLSNe-I (Tab. \ref{Tab:Magnetar}, \citealt{Nicholl13,Inserra13}).

We end by noting that our analytic treatment of ionization break-out from the supernova ejecta requires confirmation by a more detailed photo-ionization calculation in future work, as well as a more accurate model for the spectral energy distribution of the young pulsar wind nebula.  Also, by adopting a relatively low mass fraction of $X_Z \approx 0.1$ of CNO elements (which contribute most of the bound-free opacity in the keV range) we may be underestimating the ionization break-out time and thus over-estimating the associated X-ray luminosity if the true mass fraction is higher.  On the other hand, asphericity in the ejecta (e.g. along the rotation axis) would reduce the break-out time along directions of lower than average density and introduce a viewing angle dependence to the emission. An extension from an analytical 1D model (used here) to detailed multi-D formulations is indeed necessary to fully characterize the expected X-ray signature from ionization breakout, and possibly solve the current tension between the anticipated vs. observed spectral features and their evolution in the magnetar model (e.g. \citealt{Liu16}).

%%%%%%%%%%%%%%%%%%%%%%%%%%%%%%%%%%%%%%%%%%%
\section{Summary and Conclusions}
\label{Sec:Conc}
We present the results from an extensive systematic survey of X-ray emission from 26 hydrogen-stripped SLSNe in the local Universe with \emph{Swift}, Chandra and XMM. These data cover the SLSNe-I evolution from $\sim$ days until $2000$ days (rest-frame) since explosion, reaching $L_x\sim10^{40}\,\rm{erg\,s^{-1}}$. The unprecedented depth of these observations provided the deepest limits on the X-ray emission from SLSNe-I to date and enabled the detection of X-ray emission at the location of the slowly evolving SLSN-I PTF12dam.  The major results from our investigation can be summarized as follows:
\begin{itemize}
\item Superluminous X-ray emission $L_x\sim10^{45}\,\rm{erg\,s^{-1}}$ of the kind detected at the location of SCP06F6 is \emph{not} a common trait of SLSNe-I. Superluminous X-ray emission requires peculiar physical conditions that are likely not shared by the majority of SLSNe-I. \emph{If} present, its duration is $\leq$ 2 months at $t<2000$ days and  $\leq$ few days at  earlier epochs $t<100$ days (Fig. \ref{Fig:duration}).
\item We place sensitive limits on the sub-pc environments of the SLSNe-I with deepest observations and constrain  pre-explosion mass-loss history their stellar progenitors of SLSNe-I (Fig. \ref{Fig:massloss1}-\ref{Fig:massloss2}). The most sensitive X-ray observations in our sample rule out the densest environments typical of LBV eruptions and type-IIn SNe. For PTF12dam, the observations point to a clean environment similar to engine-driven SNe and argue against the dense CSM typical of extended stellar progenitors like RSG. Observations indicate $\dot M<2\times10^{-5}\,\rm{M_{\sun}yr^{-1}}$. This result suggests that CSM interaction is unlikely to play a key role in the process that powers the luminous display in some SLSNe-I and that at least some SLSN-I progenitors end their life as compact stars surrounded by a low-density medium.
\item We do not find compelling observational evidence for relativistic outflows in SLSNe-I. SLSNe-I might either be powered by energetic relativistic GRB-like outflows that we did not detect because pointed far away from our line of sight ($\theta_{obs}>30\degree$), or might harbor failed jets that do not successfully pierce through the stellar envelope. Deep X-ray observations of PTF12dam rule out even the weakest emission from uncollimated GRB outflows (Fig. \ref{Fig:GRBoff}), suggesting that \emph{if} PTF12dam is a jet-driven explosion, then the jet never successfully broke out from the surface (in close similarity to the relativistic SNe 2009bb and 2012ap). However, the SLSN-I 2011kl was found in association with the fully-relativistic, fully-successful jet of GRB111209A. We thus propose that, just like H-stripped core-collapse SN, SLSNe-I might also be characterized by a continuum of jet power and central-engine life-times.
\item The X-ray ionization break out is a very sensitive probe of the properties of a hidden magnetar central engine in SLSNe-I. Magnetar central engines that would produce very similar optical/UV displays are instead clearly differentiated in terms of X-ray luminosities and time scales of the ionization break out (Fig. \ref{Fig:14bj}-\ref{Fig:PS15aeLbol}). Current X-ray observations indicate that \emph{if} a magnetar central engine powers SLSNe-I, then it has to be either associated with a large magnetic field $B_{14}>2$ G or large ejecta mass ($M_{ej}>20,7,3\,\rm{M_{\sun}}$ for $B_{14}=0.2,0.5,1.0$ G). 
\end{itemize}

This X-ray campaign provided constraints on the sub-pc environment and properties of central engines in SLSNe-I. To further advance our knowledge and understanding of SLSNe-I it is necessary to systematically explore the region of the parameter space with $L_x<10^{41}\,\rm{erg\,s^{-1}}$ both at very early ($t<30$ days, rest-frame) and late times ($t>1000$ days, rest-frame), where the emission from an off-axis relativistic jet, weak uncollimated relativistic outflow or magnetar ionization breakout might be found.
%might live. 
This parameter space is almost an entirely uncharted territory of exploration and holds promise for future discoveries.

%%%%%%%%%%%%%%%%%%%%%%%%%%%%%%%%%%%%%%%%%%%
\bigskip
\bigskip\bigskip\bigskip
This research has made use of the
XRT Data Analysis Software (XRTDAS) developed under the responsibility
of the ASI Science Data Center (ASDC), Italy. We acknowledge the use of public data from the Swift data archive. This work is partially based  on data acquired with the Swift GO program 1114109 (grant NNX16AT51G, PI Margutti).
The scientific results reported in this article are partially based on observations made by the Chandra X-ray Observatory under program GO6-17052A, PI Margutti, observations IDs 17879, 17880, 17881, 17882 and  IDs 13772, 14444, 14446, (PI Pooley). Partially based on observations by XMM-\emph{Newton}, IDs 0743110301, 0743110701, 0770380201, 0770380401, PI Margutti, proposal 74311.
CG acknowledges University of Ferrara for use of the local HPC facility co-funded by the ``Large-Scale Facilities 2010'' project (grant 7746/2011).
Development of the Boxfit code was supported in part by NASA through grant NNX10AF62G issued through the Astrophysics Theory Program and by the NSF through grant AST-1009863. Simulations for BOXFIT version 2 have been carried out in part on the computing facilities of the Computational Center for Particle and Astrophysics (C2PAP) of the research cooperation "Excellence Cluster Universe" in Garching, Germany. G.M. acknowledges the financial support from the UnivEarthS Labex program of Sorbonne Paris Cité (ANR10LABX0023 and ANR11IDEX000502).
%\textcolor{red}{Add the grant number that you will be using to pay for publication charges}
%%%%%%%%%%%%%%%%%%%%%%%%%%%%%%%%%%%%%%%%%%%
\bibliographystyle{yahapj}
%\bibliography{margutti}

%%%%%%%%%%%%%%%%%%%%%%%%%%%%%%%%%%%%%%%%%%%
\appendix
\section{X-ray observations of SLSNe-I }
\label{App}

We provide here the details about the X-ray observations of SLSNe-I in the ``bronze'' and ``iron'' samples.  Table \ref{Tab:master} reports the measured fluxes for the entire sample of 26 SLSNe-I analyzed in this paper. For the non-detections we assume a non-thermal power-law spectrum with photon index $\Gamma=2$ and Galactic absorption.

%----------------------------------------------------------------------------------------
\subsection{SN2009jh/PTF09cwl}
\emph{Swift}-XRT observed SN2009jh \citep{Quimby11} on 2009  August 29 until 2016 Sep 25 ($\delta t=48-1961$ days rest-frame since explosion). No X-ray source is detected at the location of the supernova. With respect to \cite{Levan13} we add the late-time data-set  acquired in 2016. 
%----------------------------------------------------------------------------------------
\subsection{PTF09atu}
\emph{Swift}-XRT observed PTF09atu  \citep{Quimby11} on 2009 August 19 until  2016 October 6($\delta t=49-1785$  days rest-frame since explosion). No X-ray source is detected at the location of the supernova. With respect to \cite{Levan13} we add the late-time data-set  acquired in 2016. 
%----------------------------------------------------------------------------------------
\subsection{PTF09cnd}
\emph{Swift}-XRT started observing PTF09cnd \citep{Quimby11} on 2014 August 8 until October 3 ($\delta t=1487-1531$  days rest-frame since explosion, exposure time of 26 ks). This data set has been presented by \cite{Levan13}. The location of PTF09cnd was serendipitously observed by \emph{Swift} between 2016 February 4 and September 23 ($\delta t=1921-2104$  days rest-frame since explosion, exposure time of 62 ks). No X-ray source is detected at the location of the supernova.  
XMM observed the location of PTF09cnd on 2014 August 8 ($\delta t=1487$  days rest-frame since explosion). The net exposure time is 27.7 ks (EPIC-pn data). No source is detected and we derive  a 3$\sigma$ count-rate upper limit of $1.5\times 10^{-3}\,\rm{c\,s^{-1}}$ (0.3 -10 keV), which corresponds to an absorbed (unabsorbed) flux $<3.2\times 10^{-15}\,\rm{erg\,s^{-1}cm^{-2}}$  ($<3.4\times 10^{-15}\,\rm{erg\,s^{-1}cm^{-2}}$).
%----------------------------------------------------------------------------------------
\subsection{PTF10aagc}
\emph{Swift}-XRT observed PTF10aagc \citep{Yan15} on 2010 Nov 08  ($\delta t=79$  days rest-frame since explosion). No X-ray source is detected at the location of the supernova.
%----------------------------------------------------------------------------------------
\subsection{SN2010md/PTF10hgi}
\emph{Swift}-XRT started observing  PTF10hgi \citep{Inserra13} on 2010 July 13 until 2010 July 18 ($\delta t=61-66$  days rest-frame since explosion). No X-ray source is detected at the location of the supernova as reported by \cite{Levan13}.

%----------------------------------------------------------------------------------------
\subsection{SN2010gx/CSS100313/PTF10cwr}
\emph{Swift}-XRT started observing SN2010gx \citep{Pastorello10,Quimby11,Chen13,Inserra13,Perley16} on 2010  March 19 until 2012 May 14 ($\delta t=19-659$ days rest-frame since explosion). A portion of this data set has been presented by \cite{Levan13}. Here we add the observations acquired in 2012. No X-ray source is detected at the location of the supernova. 
%----------------------------------------------------------------------------------------
\subsection{SN2010kd}
\emph{Swift}-XRT started observing  SN2010kd \citep{Vinko10,Vinko12} on 2010  Nov 30 until 2016 June 21 ($\delta t=120-1964$ days rest-frame  since explosion). With respect to \cite{Levan13}, here we include late-time data acquired in 2014 and 2016. No X-ray source is detected at the location of the supernova.
%----------------------------------------------------------------------------------------
\subsection{SN2011ke/CSS110406/PTF11dij}
\emph{Swift}-XRT started observing  SN2011ke \citep{Quimby11b,Drake13,Perley16} on 2011 May 14 until 2012 April 12  ($\delta t=40-332$  days rest-frame since explosion), as reported by \cite{Levan13}. No X-ray source is detected at the location of the supernova. 

The CXO serendipitously imaged the sky location of SN2011ke on August 28, 2015 ($\delta t=1411$ days rest-frame since explosion, exposure time of 56 ks) and on April 4, 2016 ($\delta t=1604$  days rest-frame since explosion, exposure time of 59 ks). These data are presented here for the first time.
No X-ray source is detected at the location of SN2011ke and we infer a 3$\sigma$ count-rate upper limit $<5.4\times 10^{-5}\,\rm{c\,s^{-1}}$ and $<5.1\times 10^{-5}\,\rm{c\,s^{-1}}$  in the 0.5-8 keV energy band, for the first and second epoch, respectively. For a non-thermal power-law spectrum with index $\Gamma=2$ these results translate into unabsorbed 0.3-10 keV flux limits of $<6.4\times 10^{-16}\,\rm{erg\,s^{-1}cm^{-2}}$ and $<6.0\times 10^{-16}\,\rm{erg\,s^{-1}cm^{-2}}$. 
%----------------------------------------------------------------------------------------
\subsection{PS1-11bdn}
\emph{Swift}-XRT started observing  PS1-11bdn \citep{Lunnan14,Lunnan15,Schulze16} on 2012 January 11 until 2012 January 28 ($\delta t=28-35$ days rest-frame since explosion). No X-ray source is detected at the location of the supernova. 
%----------------------------------------------------------------------------------------
\subsection{PTF11rks/SN2011kg}
\emph{Swift}-XRT started observing  PTF11rks \citep{Quimby11c,Inserra13,Perley16} on 2011 December 30  until 2012 January 15 ($\delta t=11-25$ days rest-frame since explosion), as reported by  \cite{Levan13}. No X-ray source is detected at the location of the supernova.
%----------------------------------------------------------------------------------------
\subsection{SN2012il/PS1-12fo/CSS120121}
\emph{Swift}-XRT started observing  SN2012il \citep{Drake12e,Smartt12,Inserra13,Lunnan14} on 2012  Feb 13 until 2016 June 25 ($\delta t=44-1400$ days rest-frame since explosion). No X-ray source is detected at the location of the supernova. With respect to \cite{Levan13} we add here the 2016 data-set.
%----------------------------------------------------------------------------------------
\subsection{DES15C3hav}
\emph{Swift}-XRT started observing  DES15C3hav \citep{Challis16} on 2016 June 12 until 2016 Sep 13 ($\delta t=202-269$ days rest-frame since explosion). No X-ray source is detected at the location of the supernova.
%----------------------------------------------------------------------------------------
\subsection{iPTF13ehe}
\emph{Swift}-XRT observed iPTF13ehe (\citealt{Yan15},\citealt{Wang15}) on 2014 December 23 ($\delta t=385$  days rest-frame since explosion). No X-ray source is detected at the location of the supernova.
%----------------------------------------------------------------------------------------
\subsection{LSQ14an}
\emph{Swift}-XRT started observing LSQ14an \citep{Leget14,Jerkstrand16,Inserra17} on 2014  March 24 until 2014 December 8, with a final observation taken on 2016 August 8  ($\delta t=196-949$  days rest-frame since explosion). A portion of the data set has been presented in \cite{Inserra17}. Here we present the complete data set of X-ray observations available on LSQ14an.
%----------------------------------------------------------------------------------------
\subsection{LSQ14fxj}
\emph{Swift}-XRT started observing LSQ14fxj \citep{SmithM14,Schulze16} on 2014 October 29  until 2015 June 16 ($\delta t=64-234$ days rest-frame since explosion). No X-ray source is detected at the location of the supernova.
%----------------------------------------------------------------------------------------
\subsection{LSQ14mo}
\emph{Swift}-XRT started observing LSQ14mo \citep{Leloudas15,Chen16} on 2014 January 31, with a last observation taken on 2016 July 24 ($\delta t=52-774$ days rest frame since explosion). No X-ray source is detected at the location of the supernova.
%----------------------------------------------------------------------------------------
\subsection{CSS140925-005854}
\emph{Swift}-XRT started observing CSS140925-005854 \citep{Campbell14,Schulze16} on 2014 October 11 until 2015  May 29 ($\delta t=29-186$ days rest-frame since explosion). No X-ray source is detected at the location of the supernova.
%----------------------------------------------------------------------------------------
\subsection{DES15S2nr}
\emph{Swift}-XRT started observing DES15S2nr \citep{DAndrea15} on 2015 September 25  until 2016 February 15m with another observation acquired on 2016 Sep 14 ($\delta t=32-323$ days res-frame since explosion).  No X-ray source is detected at the location of the supernova.
%----------------------------------------------------------------------------------------
\subsection{OGLE15qz}
\emph{Swift}-XRT observed OGLE15qz \citep{Kangas15,Kostrzewa-Rutkowska15} on 2015 November 25  ($\delta t=54$ days rest-frame since explosion). No X-ray source is detected at the location of the supernova.
%----------------------------------------------------------------------------------------
\subsection{OGLE15sd}
\emph{Swift}-XRT started observing OGLE15sd \citep{Baumont15} on 2015 December 8 until 2015 December 9 ($\delta t=34-212$  days rest-frame since explosion). No X-ray source is detected at the location of the supernova.
%----------------------------------------------------------------------------------------
\subsection{PS16aqv}
\emph{Swift}-XRT started observing PS16aqv \citep{Chornock16} on 2016 March 9 until 2016 June 10 ($\delta t=53-131$ days rest-frame since explosion). No X-ray source is detected at the location of the supernova.
%----------------------------------------------------------------------------------------
\subsection{PS16op}
\emph{Swift}-XRT observed PS16op \citep{Dimitriadis16} on 2016 January 20 ($\delta t=57$ days rest-frame since explosion). No X-ray source is detected at the location of the supernova.
%----------------------------------------------------------------------------------------
\LongTables
\begin{deluxetable*}{lcccccc}[h!]
\tabletypesize{\scriptsize}
%\rotate
\tablecolumns{4} 
\tablewidth{40pc}
\tablecaption{X-ray observations of SLSNe-I}
\tablehead{  \colhead{SN} & \colhead{$t_{START}$} & \colhead{$t_{STOP}$}& Unabsorbed Flux (0.3-10 keV) & \colhead{Instrument}\\
 &  (MJD) & (MJD) & ($10^{-14}\rm{erg\,s^{-1}cm^{-2}}$) &  &}
\startdata
 %%%%%%%
 SCP06F6 & 53949& 53949 & 13. & XMM\footnote{From \cite{Levan13}.}\\
 			& 54043& 54043 & < 1.40 & Chandra \\
 %%%%%%%
PTF09atu & 55062.188 & 55062.328 & <    14.50 &Swift-XRT \\
 & 57667.055 & 57667.055 & <    89.37 &Swift-XRT \\
 %%%%%%%
PTF09cnd & 55061.883 & 55062.000 & <    12.02 &Swift-XRT \\
 & 55065.828 & 55065.945 & <    12.35 &Swift-XRT \\
 & 55069.016 & 55069.289 & <    14.47 &Swift-XRT \\
 & 55073.637 & 55073.777 & <    12.43 &Swift-XRT \\
 & 55077.312 & 55077.594 & <    17.12 &Swift-XRT \\
 & 55084.012 & 55084.887 & <    13.73 &Swift-XRT \\
 & 55097.812 & 55097.945 & <    25.94 &Swift-XRT \\
 & 55107.375 & 55107.453 & <    22.40 &Swift-XRT \\
 & 57422.016 & 57422.766 & <    57.30 &Swift-XRT \\
 & 57426.078 & 57426.406 & <    41.68 &Swift-XRT \\
 & 57435.129 & 57435.934 & <   173.40 &Swift-XRT \\
 & 57450.680 & 57450.820 & <    32.82 &Swift-XRT \\
 & 57456.250 & 57456.266 & <   167.44 &Swift-XRT \\
 & 57463.633 & 57463.977 & <    37.95 &Swift-XRT \\
 & 57520.109 & 57520.266 & <    29.88 &Swift-XRT \\
 & 57562.781 & 57562.930 & <    21.62 &Swift-XRT \\
 & 57565.641 & 57565.781 & <   104.19 &Swift-XRT \\
 & 57568.098 & 57568.309 & <    16.07 &Swift-XRT \\
 & 57575.074 & 57575.418 & <    11.76 &Swift-XRT \\
 & 57590.371 & 57590.793 & <     9.41 &Swift-XRT \\
 & 57653.570 & 57653.977 & <    10.31 &Swift-XRT \\
 & 57654.562 & 57654.906 & <    10.42 &Swift-XRT \\
 & 56877.875 & 56878.250 & <     0.34 &XMM \\
 %%%%%%%
 SN2009jh/PTF09cwl & 55072.574 & 55072.652 & <    12.88 &Swift-XRT \\
 & 57641.000 & 57641.016 & <    55.08 &Swift-XRT \\
 & 57643.477 & 57643.477 & <   211.95 &Swift-XRT \\
 & 57656.555 & 57656.570 & <    55.45 &Swift-XRT \\
 %%%%%%%
PTF10aagc & 55508.016 & 55508.094 & <    15.99 &Swift-XRT \\
%%%%%%%
SN2010gx & 55274.141 & 55274.281 & <    13.61 &Swift-XRT \\
 & 55275.672 & 55275.688 & <    62.42 &Swift-XRT \\
 & 55280.570 & 55280.648 & <    32.07 &Swift-XRT \\
 & 55286.203 & 55286.469 & <    33.74 &Swift-XRT \\
 & 55294.086 & 55294.430 & <    27.03 &Swift-XRT \\
 & 55302.512 & 55302.590 & <    31.56 &Swift-XRT \\
 & 55309.617 & 55309.633 & <    32.22 &Swift-XRT \\
 & 55318.578 & 55318.656 & <    19.97 &Swift-XRT \\
 & 56055.809 & 56055.887 & <    24.77 &Swift-XRT \\
 & 56060.750 & 56061.297 & <    35.85 &Swift-XRT \\
 %%%%%%%
SN2010kd & 55530.133 & 55530.195 & <    32.89 &Swift-XRT \\
 & 55530.914 & 55530.930 & <    45.20 &Swift-XRT \\
 & 55546.297 & 55546.383 & <    13.49 &Swift-XRT \\
 & 55549.047 & 55549.133 & <    18.50 &Swift-XRT \\
 & 55552.387 & 55552.535 & <    14.20 &Swift-XRT \\
 & 55555.812 & 55555.953 & <    12.21 &Swift-XRT \\
 & 56980.895 & 56980.910 & <    67.81 &Swift-XRT \\
 & 56982.914 & 56982.914 & <   187.60 &Swift-XRT \\
 & 56985.430 & 56985.906 & <    28.37 &Swift-XRT \\
 & 56987.027 & 56988.973 & <    29.46 &Swift-XRT \\
 & 57388.781 & 57388.984 & <    42.94 &Swift-XRT \\
 & 57392.109 & 57392.781 & <    22.15 &Swift-XRT \\
 & 57396.367 & 57396.836 & <    20.91 &Swift-XRT \\
 & 57413.969 & 57413.969 & <   207.18 &Swift-XRT \\
 & 57469.016 & 57470.000 & <    18.83 &Swift-XRT \\
 & 57475.844 & 57475.844 & <   204.92 &Swift-XRT \\
 & 57478.109 & 57478.109 & <   247.12 &Swift-XRT \\
 & 57479.164 & 57479.164 & <   171.40 &Swift-XRT \\
 & 57489.023 & 57489.820 & <   125.06 &Swift-XRT \\
 & 57497.000 & 57497.000 & <    93.42 &Swift-XRT \\
 & 57498.590 & 57498.590 & <   220.37 &Swift-XRT \\
 & 57523.902 & 57523.910 & <   103.97 &Swift-XRT \\
 & 57525.906 & 57525.906 & <   168.76 &Swift-XRT \\
 & 57533.000 & 57533.000 & <   260.30 &Swift-XRT \\
 & 57560.129 & 57560.996 & <    21.92 &Swift-XRT \\
 %%%%%%%
SN2010md/PTF10hgi & 55390.754 & 55390.824 & <    28.96 &Swift-XRT \\
 & 55395.219 & 55395.297 & <    16.37 &Swift-XRT \\
%%%%%%%
SN2011ke & 55695.234 & 55695.312 & <    19.56 &Swift-XRT \\
 & 55711.633 & 55712.039 & <     9.83 &Swift-XRT \\
 & 55718.668 & 55718.738 & <    40.49 &Swift-XRT \\
 & 55719.121 & 55719.887 & <    22.74 &Swift-XRT \\
 & 55720.195 & 55720.211 & <    28.12 &Swift-XRT \\
 & 55994.992 & 55994.992 & <  1106.19 &Swift-XRT \\
 & 55997.070 & 55998.070 & <    46.63 &Swift-XRT \\
 & 56000.152 & 56000.746 & <   110.62 &Swift-XRT \\
 & 56004.945 & 56004.961 & <    57.84 &Swift-XRT \\
 & 56008.445 & 56008.445 & <   118.21 &Swift-XRT \\
 & 56017.109 & 56017.117 & <   104.83 &Swift-XRT \\
 & 56020.449 & 56020.457 & <    88.93 &Swift-XRT \\
 & 56029.555 & 56029.555 & <   354.63 &Swift-XRT \\
 & 57481.375 & 57482.078 & <     0.06 &Chandra \\
 & 57262.688 & 57263.375 & <     0.06 &Chandra \\
 %%%%%%%
PS1-11bdn & 55937.477 & 55937.758 & <     8.46 &Swift-XRT \\
 & 55950.359 & 55950.656 & <    11.36 &Swift-XRT \\
 & 55953.258 & 55954.320 & <    18.02 &Swift-XRT \\
 %%%%%%%
PTF11rks & 55925.223 & 55925.301 & <    34.22 &Swift-XRT \\
 & 55927.055 & 55927.367 & <    11.94 &Swift-XRT \\
 & 55931.312 & 55931.938 & <    15.17 &Swift-XRT \\
 & 55936.203 & 55936.750 & <     8.83 &Swift-XRT \\
 & 55941.000 & 55941.562 & <    11.41 &Swift-XRT \\
 %%%%%%%
PTF12dam & 56069.836 & 56069.977 & <    17.56 &Swift-XRT \\
 & 56077.062 & 56077.547 & <    20.73 &Swift-XRT \\
 & 56085.156 & 56085.234 & <    43.10 &Swift-XRT \\
 & 56091.172 & 56091.188 & <    25.54 &Swift-XRT \\
 & 56098.004 & 56098.066 & <    23.20 &Swift-XRT \\
 & 56105.398 & 56105.742 & <    44.54 &Swift-XRT \\
 & 56106.930 & 56106.945 & <    45.62 &Swift-XRT \\
 & 56112.145 & 56112.223 & <    52.08 &Swift-XRT \\
 & 56119.703 & 56119.781 & <    16.38 &Swift-XRT \\
 & 56126.320 & 56126.336 & <    38.08 &Swift-XRT \\
 & 56990.820 & 56990.836 & <    28.84 &Swift-XRT \\
 & 57006.137 & 57006.473 & <    39.15 &Swift-XRT \\
 & 57008.203 & 57008.812 & <    43.46 &Swift-XRT \\
 & 57017.992 & 57018.266 & <    35.63 &Swift-XRT \\
 & 56089.832 & 56097.293 &     0.07$\pm$0.03 &Chandra \\
 %%%%%%%
SN2012il/PS1-12fo & 55970.703 & 55970.734 & <    23.48 &Swift-XRT \\
 & 55970.992 & 55971.133 & <    30.80 &Swift-XRT \\
 & 57396.363 & 57396.363 & <   343.90 &Swift-XRT \\
 & 57442.691 & 57442.887 & <   102.68 &Swift-XRT \\
 & 57443.023 & 57443.023 & <   339.37 &Swift-XRT \\
 & 57444.270 & 57444.270 & <   562.48 &Swift-XRT \\
 & 57461.414 & 57461.414 & <   777.65 &Swift-XRT \\
 & 57469.656 & 57469.656 & <  1468.47 &Swift-XRT \\
 & 57478.430 & 57478.430 & <   278.60 &Swift-XRT \\
 & 57479.094 & 57480.094 & <    91.73 &Swift-XRT \\
 & 57519.977 & 57519.977 & <   426.58 &Swift-XRT \\
 & 57531.602 & 57531.875 & <    26.46 &Swift-XRT \\
 & 57558.855 & 57559.793 & <    26.16 &Swift-XRT \\
 & 57563.703 & 57564.641 & <    30.80 &Swift-XRT \\
 %%%%%%%
iPTF13ehe & 57014.238 & 57014.832 & <     8.61 &Swift-XRT \\
%%%%%%%
CSS140925-005854 & 56941.695 & 56941.914 & <     9.72 &Swift-XRT \\
 & 56942.289 & 56942.438 & <    15.10 &Swift-XRT \\
 & 56995.039 & 56995.930 & <     6.30 &Swift-XRT \\
 & 57032.059 & 57032.480 & <     5.98 &Swift-XRT \\
 & 57171.008 & 57171.484 & <     7.09 &Swift-XRT \\
%%%%%%%
 LSQ14an & 56740.516 & 56740.938 & <     8.13 &Swift-XRT \\
 & 56839.305 & 56839.570 & <    39.80 &Swift-XRT \\
 & 56841.031 & 56841.906 & <     9.50 &Swift-XRT \\
 & 56842.172 & 56842.781 & <    26.84 &Swift-XRT \\
 & 56846.840 & 56846.848 & <    65.91 &Swift-XRT \\
 & 56862.039 & 56862.039 & <   156.22 &Swift-XRT \\
 & 56868.172 & 56868.172 & <   141.67 &Swift-XRT \\
 & 56998.742 & 56999.211 & <     6.21 &Swift-XRT \\
 & 57616.012 & 57616.488 & <     6.55 &Swift-XRT \\
%%%%%%%
LSQ14fxj & 56959.477 & 56959.492 & <    47.59 &Swift-XRT \\
 & 56960.156 & 56960.172 & <    70.61 &Swift-XRT \\
 & 56962.812 & 56962.828 & <    39.40 &Swift-XRT \\
 & 56964.539 & 56964.555 & <    28.48 &Swift-XRT \\
 & 56966.070 & 56966.086 & <    42.13 &Swift-XRT \\
 & 56997.109 & 56998.000 & <     5.27 &Swift-XRT \\
 & 57033.000 & 57034.406 & <    10.73 &Swift-XRT \\
 & 57036.523 & 57036.930 & <    16.23 &Swift-XRT \\
 & 57039.527 & 57039.668 & <    61.25 &Swift-XRT \\
 & 57051.969 & 57051.969 & <   440.81 &Swift-XRT \\
 & 57055.383 & 57055.383 & <   252.78 &Swift-XRT \\
 & 57189.172 & 57189.992 & <     4.76 &Swift-XRT \\
%%%%%%%
LSQ14mo & 56688.758 & 56690.039 & <    32.72 &Swift-XRT \\
 & 56692.023 & 56692.039 & <    53.38 &Swift-XRT \\
 & 56693.039 & 56693.055 & <    55.55 &Swift-XRT \\
 & 56697.961 & 56697.977 & <    49.91 &Swift-XRT \\
 & 56699.711 & 56699.727 & <    63.80 &Swift-XRT \\
 & 56754.113 & 56754.676 & <     4.51 &Swift-XRT \\
 & 56836.836 & 56836.852 & <    57.29 &Swift-XRT \\
 & 56837.098 & 56838.371 & <    14.19 &Swift-XRT \\
 & 56839.023 & 56839.977 & <    18.18 &Swift-XRT \\
 & 56846.293 & 56846.371 & <    22.74 &Swift-XRT \\
 & 56953.039 & 56953.586 & <    19.53 &Swift-XRT \\
 & 57052.594 & 57054.000 & <     8.59 &Swift-XRT \\
 & 57054.398 & 57055.742 & <    22.96 &Swift-XRT \\
 & 57063.781 & 57063.789 & <    78.55 &Swift-XRT \\
 & 57586.496 & 57586.496 & <   442.68 &Swift-XRT \\
 & 57593.207 & 57593.355 & <    22.96 &Swift-XRT \\
%%%%%%%
 PS1-14bj & 56817.145 & 56817.590 & <     1.96 &XMM \\
 & 56968.219 & 56968.625 & <     0.33 &XMM \\
%%%%%%%
DES15C3hav & 57551.984 & 57552.453 & <     7.81 &Swift-XRT \\
 & 57555.062 & 57555.391 & <    22.96 &Swift-XRT \\
 & 57644.625 & 57644.773 & <    11.87 &Swift-XRT \\
  %%%%%%%
DES15S2nr & 57290.133 & 57290.148 & <    33.39 &Swift-XRT \\
 & 57290.555 & 57290.555 & <   111.70 &Swift-XRT \\
 & 57291.328 & 57291.406 & <    15.69 &Swift-XRT \\
 & 57293.461 & 57294.000 & <    12.48 &Swift-XRT \\
 & 57298.133 & 57299.000 & <    11.71 &Swift-XRT \\
 & 57303.703 & 57303.984 & <    12.68 &Swift-XRT \\
 & 57308.035 & 57308.691 & <     7.27 &Swift-XRT \\
 & 57334.172 & 57334.172 & <   337.41 &Swift-XRT \\
 & 57336.359 & 57336.969 & <     4.83 &Swift-XRT \\
 & 57433.484 & 57433.953 & <     7.65 &Swift-XRT \\
 & 57645.023 & 57645.227 & <     9.82 &Swift-XRT \\
%%%%%%%
 OGLE15qz & 57351.859 & 57351.922 & <    62.99 &Swift-XRT \\
%%%%%%%
OGLE15sd & 57364.891 & 57364.930 & <   619.08 &Swift-XRT \\
 & 57364.953 & 57365.062 & <    56.75 &Swift-XRT \\
%%%%%%%
SN2015bn/PS15ae & 57072.289 & 57072.508 & <     6.68 &Swift-XRT \\
 & 57073.562 & 57074.781 & <    13.44 &Swift-XRT \\
 & 57075.352 & 57076.977 & <    65.29 &Swift-XRT \\
 & 57078.289 & 57078.367 & <    27.40 &Swift-XRT \\
 & 57080.484 & 57080.562 & <    31.44 &Swift-XRT \\
 & 57082.277 & 57083.957 & <    32.08 &Swift-XRT \\
 & 57084.344 & 57084.359 & <    25.66 &Swift-XRT \\
 & 57089.938 & 57090.016 & <    30.98 &Swift-XRT \\
 & 57090.402 & 57090.551 & <    15.02 &Swift-XRT \\
 & 57093.594 & 57093.938 & <    25.59 &Swift-XRT \\
 & 57098.984 & 57099.000 & <    47.93 &Swift-XRT \\
 & 57100.984 & 57100.984 & <  2411.59 &Swift-XRT \\
 & 57102.578 & 57102.719 & <    30.57 &Swift-XRT \\
 & 57106.836 & 57106.914 & <    35.63 &Swift-XRT \\
 & 57110.695 & 57110.695 & <   110.20 &Swift-XRT \\
 & 57111.430 & 57111.578 & <    15.17 &Swift-XRT \\
 & 57118.352 & 57118.500 & <    30.49 &Swift-XRT \\
 & 57120.750 & 57120.891 & <    22.87 &Swift-XRT \\
 & 57122.008 & 57122.148 & <    48.68 &Swift-XRT \\
 & 57122.281 & 57123.289 & <    14.57 &Swift-XRT \\
 & 57128.594 & 57128.672 & <    16.11 &Swift-XRT \\
 & 57131.254 & 57131.801 & <    43.78 &Swift-XRT \\
 & 57131.258 & 57131.812 & <     8.57 &Swift-XRT \\
 & 57136.312 & 57136.320 & <    48.68 &Swift-XRT \\
 & 57144.305 & 57144.648 & <    14.87 &Swift-XRT \\
 & 57151.152 & 57151.230 & <    13.93 &Swift-XRT \\
 & 57160.594 & 57160.672 & <    18.12 &Swift-XRT \\
 & 57166.250 & 57166.453 & <    15.47 &Swift-XRT \\
 & 57172.375 & 57172.453 & <    19.21 &Swift-XRT \\
 & 57185.141 & 57185.281 & <    12.00 &Swift-XRT \\
 & 57206.758 & 57206.906 & <    17.66 &Swift-XRT \\
 & 57231.367 & 57231.430 & <    26.15 &Swift-XRT \\
 & 57232.297 & 57232.453 & <    10.98 &Swift-XRT \\
 & 57330.652 & 57330.793 & <    28.42 &Swift-XRT \\
 & 57335.781 & 57335.797 & <    28.76 &Swift-XRT \\
 & 57357.648 & 57357.805 & <    20.15 &Swift-XRT \\
 & 57390.031 & 57390.648 & <     6.19 &Swift-XRT \\
 & 57449.070 & 57449.930 & <    17.47 &Swift-XRT \\
 & 57450.320 & 57450.867 & <    30.42 &Swift-XRT \\
 & 57455.504 & 57455.574 & <    58.12 &Swift-XRT \\
 & 57456.766 & 57456.766 & <   223.42 &Swift-XRT \\
 & 57457.359 & 57457.375 & <    38.87 &Swift-XRT \\
 & 57572.016 & 57572.086 & <    28.15 &Swift-XRT \\
 & 57573.215 & 57574.215 & <    11.06 &Swift-XRT \\
 & 57576.141 & 57576.336 & <    75.48 &Swift-XRT \\
 & 57580.852 & 57580.930 & <    26.38 &Swift-XRT \\
 & 57582.246 & 57582.715 & <    23.89 &Swift-XRT \\
 & 57584.633 & 57584.836 & <    30.38 &Swift-XRT \\
 & 57586.633 & 57586.836 & <    29.63 &Swift-XRT \\
 & 57588.227 & 57588.227 & <    95.10 &Swift-XRT \\
 & 57590.023 & 57590.227 & <    28.23 &Swift-XRT \\
 & 57591.348 & 57592.355 & <    19.17 &Swift-XRT \\
 & 57174.250 & 57174.547 & <     0.98 &XMM \\
 & 57374.953 & 57375.250 & <     0.53 &XMM \\
 %%%%%%%
 %PS16aqv & 57456.320 & 57456.523 & <     8.53 &Swift-XRT \\
 %& 57460.578 & 57460.922 & <    14.08 &Swift-XRT \\
 %& 57464.559 & 57464.965 & <    10.55 &Swift-XRT \\
 %& 57467.953 & 57468.891 & <     9.40 &Swift-XRT \\
 %& 57471.016 & 57471.141 & <    21.42 &Swift-XRT \\
 %& 57476.062 & 57476.992 & <    13.37 &Swift-XRT \\
 %& 57477.109 & 57477.797 & <    37.29 &Swift-XRT \\
 %& 57480.500 & 57480.719 & <    10.08 &Swift-XRT \\
 %& 57484.621 & 57484.902 & <     9.04 &Swift-XRT \\
 %& 57488.027 & 57488.895 & <    16.34 &Swift-XRT \\
% & 57492.609 & 57492.688 & <    30.15 &Swift-XRT \\
% & 57493.328 & 57493.406 & <    29.75 &Swift-XRT \\
% & 57496.188 & 57496.531 & <    24.71 &Swift-XRT \\
% & 57497.250 & 57497.672 & <    29.08 &Swift-XRT \\
% & 57548.047 & 57548.695 & <    22.61 &Swift-XRT \\
% & 57549.555 & 57549.711 & <    16.82 &Swift-XRT \\
 %%%%%%%
 PS16op & 57407.062 & 57407.594 & <    13.55 &Swift-XRT \\
 %%%%%%%
 SN2016ard/PS16aqv & 57456.320 & 57456.523 & <     8.53 &Swift-XRT \\
 & 57460.578 & 57460.922 & <    14.08 &Swift-XRT \\
 & 57464.559 & 57464.965 & <    10.55 &Swift-XRT \\
 & 57467.953 & 57468.891 & <     9.40 &Swift-XRT \\
 & 57471.016 & 57471.141 & <    21.42 &Swift-XRT \\
 & 57476.062 & 57476.992 & <    13.37 &Swift-XRT \\
 & 57477.109 & 57477.797 & <    37.29 &Swift-XRT \\
 & 57480.500 & 57480.719 & <    10.08 &Swift-XRT \\
 & 57484.621 & 57484.902 & <     9.04 &Swift-XRT \\
 & 57488.027 & 57488.895 & <    16.34 &Swift-XRT \\
 & 57492.609 & 57492.688 & <    30.15 &Swift-XRT \\
 & 57493.328 & 57493.406 & <    29.75 &Swift-XRT \\
 & 57496.188 & 57496.531 & <    24.71 &Swift-XRT \\
 & 57497.250 & 57497.672 & <    29.08 &Swift-XRT \\
 & 57548.047 & 57548.695 & <    22.61 &Swift-XRT \\
 & 57549.555 & 57549.711 & <    16.82 &Swift-XRT \\
 %%%%%%%
 \enddata
\label{Tab:master}
\end{deluxetable*}

\end{document}